\documentclass[useAMS,usenatbib, twocolumn,fleqn]{mnras}
\usepackage{times,psfig,epsfig} 
\usepackage{rotating, graphicx, adjustbox}
\usepackage{color}
\usepackage{amssymb, amsmath}
\usepackage[T1]{fontenc} 
\usepackage{ae,aecompl}
\usepackage{gensymb}
\usepackage{placeins}
\usepackage{multirow}
\usepackage{bigstrut}
\usepackage{booktabs}

\def\mvir{M_{\rm vir}}
\def\rvir{R_{\rm vir}}

\def\rfive{R_{\rm 500}}

\def\msun{\rm{\,M_{\odot}}}
\def\msunh{\,h^{-1}\rm{\,M_{\odot}}}

\def\kpch{\,h^{-1}\rm{\,kpc}}

\def\mpch{\,h^{-1}\rm{\,Mpc}}

\newcommand{\chandra}{{\it Chandra}}
\newcommand{\xmm}{{\it XMM-Newton}}
\newcommand{\suzaku}{{\it Suzaku}}
\newcommand{\ska}{{\it SKA}}

\newcommand{\mincir}{\raise
  -2.truept\hbox{\rlap{\hbox{$\sim$}}\raise5.truept \hbox{$<$}\ }}
\newcommand{\magcir}{\raise
  -2.truept\hbox{\rlap{\hbox{$\sim$}}\raise5.truept \hbox{$>$}\ }}
\newcommand{\siml}{\raise 
  -2.truept\hbox{\rlap{\hbox{$\sim$}}\raise5.truept \hbox{$<$}\ }}
\newcommand{\simg}{\raise
  -2.truept\hbox{\rlap{\hbox{$\sim$}}\raise5.truept \hbox{$>$}\ }}

\newcommand{\nr}{NR}
\newcommand{\csf}{CSF}
\newcommand{\agn}{AGN}

\title[Shock waves and galaxy clusters]
{Exploring the role of cosmological shock waves in the Dianoga simulations of galaxy clusters}

\author[S. Planelles et al.]
{S. Planelles$^{1,6}$\thanks{e-mail: susana.planelles@uv.es}, S. Borgani$^{2,3,4,5}$, V. Quilis$^{1,6}$, G. Murante$^{2,5}$, V. Biffi$^{5,7}$, E. Rasia$^{2,5}$,
\newauthor
K. Dolag$^{7,8}$, G. L. Granato$^{2,5,9}$, C. Ragone-Figueroa$^{2,9}$
\\~\\
\footnotesize 
$^1$ Departament d'Astronomia i Astrof{\'i}sica, Universitat de Val\`encia, c/ Dr. Moliner, 50, 46100 - Burjassot (Valencia), Spain\\
$^2$ INAF, Osservatorio Astronomico di Trieste, via Tiepolo 11, I-34131, Trieste, Italy \\
$^3$ Dipartimento di Fisica dell' Universit\`a di Trieste, Sezione di Astronomia, via Tiepolo 11, I-34131 Trieste, Italy \\
$^4$ INFN, Instituto Nazionale di Fisica Nucleare, Via Valerio 2, I-34127, Trieste, Italy \\
$^5$ IFPU - Institute for Fundamental Physics of the Universe, Via Beirut 2, 34014 Trieste, Italy \\
$^6$ Observatori Astron\`omic, Universitat de Val\`encia, E-46980 Paterna (Valencia), Spain\\
$^7$ University Observatory Munich, Scheinerstr. 1, 81679 Munich, Germany\\
$^8$ Max-Planck-Institut f\"ur Astrophysik, Karl-Schwarzschild Strasse 1, 85748 Garching bei M\"unchen, Germany \\
$^9$ Instituto de Astronom\'ia Te\'orica y Experimental (IATE), Consejo Nacional de Investigaciones Cient\'ificas y T\'ecnicas de la \\
Rep\'ublica Argentina (CONICET), Universidad Nacional de C\'ordoba, Laprida 854, X5000BGR, C\'ordoba, Argentina
}

\begin{document}
\maketitle 

\begin{abstract}
Cosmological shock waves are ubiquitous to cosmic structure formation and evolution. As a consequence, they play a major role in the energy distribution and thermalization of the intergalactic medium (IGM). We analyze the Mach number distribution in the Dianoga simulations of galaxy clusters performed with the SPH code GADGET-3. The simulations include the effects of radiative cooling, star formation, metal enrichment, supernova and active galactic nuclei feedback. A grid-based shock-finding algorithm is applied in post-processing to the outputs of the simulations. This procedure allows us  to explore in detail the distribution of shocked cells and their strengths as a function of cluster mass, redshift and baryonic physics. We also  pay special attention to the connection between shock waves and the cool-core/non-cool core (CC/NCC) state and the global dynamical status of the simulated clusters. In terms of general shock statistics, we obtain a broad agreement with previous works, with weak (low-Mach number) shocks filling most of the volume and processing most of the total thermal energy flux. As a function of cluster mass, we find that massive clusters seem more efficient in thermalising the IGM and tend to show larger external accretion shocks than less massive systems.  We do not find any relevant difference between CC and NCC  clusters. However, we find a mild dependence of the radial distribution of the shock Mach number on the cluster dynamical state, with disturbed systems showing stronger shocks than regular ones throughout the cluster volume. 
\end{abstract}

\begin{keywords}  
cosmology:  methods: numerical -- galaxies: cluster: general -- X-ray: galaxies. 
\end{keywords}

\section{Introduction}
\label{sec:intro}

Cosmological shock waves play a crucial role on the formation, evolution and thermalization  of the large-scale structure (LSS) in the Universe \citep[see][for reviews]{Bykov_2008, Dolag_2008, Rephaeli_2008}. Shocks are developed throughout the intergalactic medium (IGM) as a consequence of the hierarchical cosmic evolution. In a first phase of the evolution, the gravitational energy associated with the collapse of dark and baryonic matter  is converted into kinetic energy. Most part of the gas kinetic energy gets dissipated and thermalised by shocks, thereby heating the IGM as a whole and, particularly, the intracluster medium (ICM) in virialized haloes. In the subsequent phases, the development and evolution of cosmic structures, through mergers and accretion phenomena, also produce shock waves, turbulence and bulk gas motions which alter the energetic balance of the intergalactic gas. Moreover, feedback associated to smaller-scale processes such as star formation,  stellar winds and supernova (SNe) explosions, or relativistic jets from active galactic nuclei (AGN) can also create supersonic gas motions inside collapsed dark matter haloes. These hydrodynamical shocks are also responsible for transforming part of the mechanical energy released by feedback processes into thermal energy in the surrounding gas. 
In addition, besides the amplification of magnetic fields, these shocks, both on cosmological and galactic scales, can also accelerate particles by means of the diffusive shock acceleration mechanism (DSA; e.g. \citet{Axford_1977, Blandford_1978}; see also \citet{Blasi_2007} for a review), producing a population of relativistic cosmic rays (CR) that may significantly affect gas observable properties.

Observing and studying shocks is of fundamental importance to improve our understanding of the thermalization and energetics of the LSS.
Although shocks can be observed at different wavelengths, their detection is challenging. In X-ray observations the standard approach is to look for jumps in the ICM thermal quantities, such as gas density or temperature.
With \chandra, for instance, the standard procedure is to look for jumps in surface brightness and then obtain the temperature from the spectra extracted from the regions before and after the jump.
Pressure maps might be done in order to confirm what already known from gas density and temperature.
However, since most of the X-ray emission comes from central cluster regions, where both gas density and temperature are high, there is not much difference between shocked and unshocked gas, making quite difficult the detection of shocks. 
Still, thanks to \xmm, \chandra\ and \suzaku\ observations, a number of merging shocks with relatively low Mach numbers ($\mathcal{M}\sim1.5-3$) have been confirmed in several clusters \citep[e.g.][]{Markevitch_2002,Markevitch_2005,Dasadia_2016}. By means of {high-resolution observations of} the thermal Sunyaev-Zel'dovich effect \citep[][]{SZ_1980}, pressure jumps connected to shocks have been also measured  in the Coma cluster \citep[$\mathcal{M}\sim 2$;][]{Planck_2013_X,Erler_2015}.
ICM shocks can also be detected at radio wavelengths. In this energy band CR electrons can be accelerated or reaccelerated by merger and accretion shocks, producing synchrotron emission which tends to be located close to the accelerating shock \citep[see, e.g.][for reviews]{Bruggen_2012,Brunetti_2014}. In this sense, there is a clear connection between observed radio relics, usually identified in outer cluster regions, and merger shocks \citep[e.g.][]{Ferrari_2008}. 

The formation and evolution of  shock waves have also been extensively studied by means of cosmological simulations, either grid-based \citep[e.g.][]{Quilis1998, Miniati_2001b, Ryu_2003, Kang_2007, Skillman_2008, Vazza_2009, Vazza_2010, Vazza_2011b, Planelles2013, Sergio_2017}, particle-based \citep[e.g.][]{Pfrommer_2006, Hoeft_2008, Vazza_2011b,Zhang_2020} or moving-mesh simulations \citep[][]{Schaal_2015,Schaal_2016}. While most of these studies are based on non-radiative simulations \citep[e.g.][]{Quilis1998, Miniati_2001b, Ryu_2003, Pfrommer_2006, Hoeft_2008, Skillman_2008, Vazza_2009, Vazza_2010, Vazza_2011b, Schaal_2015, Ha_2018}, only a few of them are based on simulations including radiative cooling and star formation \citep[e.g.][]{Kang_2007,Pfrommer_2007, Planelles2013, Hong_2014,Hong_2015, Sergio_2017} or AGN feedback \citep[e.g.][]{Vazza_2013, Vazza_2014, Schaal_2016}.
Despite the diversity of cosmological codes and the different shock-finding methodologies, these numerical works report a reasonable agreement on  some general shock properties and on their distribution. In this sense, simulations agree in that shock waves can be broadly divided in internal and external shocks. 
In general, weak shocks, with $\mathcal{M}\sim 2-3$, are developed inside collapsed structures and play a crucial role in the thermalization process of haloes; for this reason they are often referred to as `internal' shocks in the literature\footnote{Although it is not strictly correct, it is quite common in the literature to indistinctly refer to internal (external) shocks as weak (strong) shocks, since most of the shocks developed within collapse structures have lower Mach numbers than external ones.}. Indeed, these internal shocks are the main responsible for the energy dissipation in the Universe. On the other hand,  the thermal history of galaxy clusters at large scales is controlled by the accretion of material onto dark matter potential wells and the exchange of gravitational energy into gas thermal energy. This process takes place through the heating of the gas by stronger (external) accretion shocks, with $\mathcal{M}\sim 10-100$, developed in the outskirts of clusters and around filaments. 
Given that the position of these accretion shocks results from the collision of internal and external shocks \citep{Zhang_2020}, both kind of shocks contribute to the total kinetic energy deposited in the medium.
Numerical simulations, however, tend to produce significant differences in some quantities related to the injection of CRs by DSA mechanisms, limiting our current understanding on this process. In this regard, although the dynamical contribution of CRs is usually addressed in post-processing, there have been some attempts to self-consistently include this contribution in cosmological simulations \citep[e.g.][]{Pfrommer_2006,Pfrommer_2007,Pfrommer_2008,Vazza_2016b}. Besides, some authors have also reported some divergences in the shock patterns when comparing results from grid-based  and particle-based simulations \citep[e.g.][]{Vazza_2011b}.

Therefore, a current challenge for present cosmological simulations is to be able to, consistently, understand and simulate the complex interaction between gravitational and astrophysical processes at play in the formation and evolution of  cosmic structures, from the smallest to the largest cosmological scales \citep[e.g.][for a review]{Planelles2015}. In this work we analyze the distribution of shocks, mainly accretion shocks, in the Dianoga simulations \citep[see also][]{Bassini_2020}, a set of zoom-in cosmological hydrodynamical simulations of galaxy clusters performed with the SPH code GADGET-3. These simulations account for the effects of radiative cooling, star formation, SNe and AGN feedback. The results obtained from the analysis of these simulations have shown a satisfactory agreement with a number of cluster observations, such as the X-ray and SZ scaling relations \citep{Truong_2018,Planelles_2017}, the distribution of metals \citep[e.g.][]{Biffi_2017, Biffi_2018, Truong_2019}, or the thermal and chemodynamical properties of cool-core (CC) and non-cool-core (NCC) clusters \citep{Rasia_2015}. Specifically, we will employ these simulations to explore in detail the distribution of shock strengths as a function of cluster mass, redshift and feedback processes. We will also  pay special attention to the connection between the Mach number distribution and the dynamical and cool-coreness state of clusters.
  
The paper is organised as follows. In Section \ref{sec:codes} we describe the numerical details of this analysis, i.e., the main properties of the Dianoga simulations analyzed in this work, the shock-finding algorithm employed, and the specific numerical procedure we have followed in this study. Section \ref{sec:results} reports the results obtained on the global Mach number distribution in our simulations and on the connection between cluster properties and cosmic shock waves. Finally, we summarise and discuss our findings in Section \ref{sec:conclusions}. 
For the sake of completeness, in Appendix \ref{app:resol}  we discuss the dependence of our results on the grid resolution, in  Appendix \ref{app:clusters_and_shocks} we analyze in more detail the connection between the distribution of shock waves and the dynamical and cool-coreness state of the clusters and, finally, in Appendix \ref{app:physics} we explore the dependence of our results on the baryonic physics included in the simulations.

\section{Numerical details}
\label{sec:codes}

In this Section we briefly introduce the main  properties of the cosmological simulations to be analyzed, a brief description of the employed shock-finding algorithm, and the details of the particular numerical procedure adopted to obtain the distribution of shock waves.    

\subsection{Cosmological simulations and cluster sample}
\label{sec:simu}

The Dianoga set of simulations analyzed in this work has been performed with an improved version of the TreePM-SPH code GADGET-3 \citep{Springel_2005}. This  version of the code includes an updated hydrodynamical scheme  that has been shown to ameliorate the performance of standard SPH algorithms in a number of issues \citep[see][for details]{Beck_2015}. Here we only provide a brief description of the simulations, while we defer the interested reader to previous works where different aspects of the same set of simulations were investigated \citep{Rasia_2015, Truong_2018, Villaescusa_2016, Planelles_2017, Biffi_2017, Biffi_2018, Ragone_2018, Truong_2019}.

The set of simulations consists in re-simulations of 29 Lagrangian regions extracted from a larger parent dark matter (DM) only simulation \citep[see][for details on the initial conditions]{Bonafede2011}. The 29 regions were identified at $z=0$ around 29 massive dark matter haloes  with $M_{200}$\footnote{$M_{\Delta}$ is the mass within $R_{\Delta}$, which is the radius of a sphere enclosing an average density equal to $\Delta$ times the critical cosmic density, $\rho_c(z)$. In the following, we will refer to the virial  overdensity \citep{Bryan1998}, that is $\Delta_{vir} \simeq 93$ at $z=0$ for our cosmology,  or to $\Delta = 180, 200, 500, 2500$.}$\sim1-30\times10^{14}h^{-1}\,M_{\odot}$.  
Each of the low-resolution regions has been re-simulated at a finer resolution
and by incorporating the baryonic component \citep[see][for details]{Planelles_2013_b}. 
The simulations are based on a flat $\Lambda$CDM cosmology with
$\Omega_{\rm{m}} = 0.24$, $\Omega_{\rm{b}} = 0.04$,
$H_0=$~72~km~s$^{-1}$~Mpc$^{-1}$,  $\sigma_8 =0.8$, and $n_{\rm{s}}=0.96$.
For the Dianoga simulations analyzed in this paper, the mass resolution for the DM (gas) particles is $m_{\rm{DM}} =
8.44\times10^8 \, \msun$ ($m_{\rm{gas}} = 1.53\times10^8\, \msun$). 
In the high-resolution region the gravitational force
is computed with a Plummer-equivalent softening length of $\epsilon =
3.75\, h^{-1}$ kpc for DM and gas   
and $\epsilon = 2\, h^{-1}$ kpc for black hole and stellar particles. 

These simulations have been run with three different prescriptions for the physics of baryons.
The most complete version of these simulations, labelled as {\tt \agn}, accounts for the effects of a number of gas physical processes such as  gas radiative cooling, star formation, SNe feedback, metal enrichment and a novel AGN feedback scheme that accounts for both cold and hot gas accretion onto supermassive black holes \citep[][]{Steinborn_2015}. There are two additional sets of simulations that only differ with the {\tt \agn} runs in the baryonic physics they include: the {\tt \nr} runs, that only account for non-radiative physics,  and the {\tt \csf} runs, that are like the {\tt \agn} ones but without including the effect of AGN feedback.  

In the following, unless otherwise specified, we will consider the {\tt \agn}  simulation as our reference set. 
The set of {\tt \agn}  cosmological simulations has been shown to be extremely successful in reproducing some of the main observational cluster properties. These simulations have produced, for the first time and for a statistically significant cluster sample, the CC/NCC dichotomy according to thermal and chemical cluster core properties \citep[see][]{Rasia_2015}. They have also shown an excellent agreement with cluster observations in  terms of the X-ray and SZ scaling relations  \citep[both at low and high redshift;][]{Planelles_2017, Truong_2018}, {the amount of hosted neutral hydrogen} \citep[][]{Villaescusa_2016}, the distribution of thermodynamical properties such as entropy, thermal pressure and temperature \citep{Rasia_2015, Planelles_2017}, the level of mass bias and hydrostatic equilibrium \citep{Biffi_2016}, or the ICM chemical enrichment distribution \citep{Biffi_2017,Biffi_2018,Truong_2019}.

Overall, the {\tt \agn}  simulation comprises a sample of $\sim 100$ clusters and groups with $M_{500}>3\times 10^{13}h^{-1}M_\odot$ at $z=0$. However, in this work we will only consider the sample of the main 29 central haloes at $z=0$, that is,  24 clusters with 
$M_{200}>8\times 10^{14}h^{-1}M_\odot$ and 5 isolated groups with $M_{200}$ in the range $1-4\times 10^{14}h^{-1}M_\odot$.

As described in \cite{Rasia_2015}, according to the core thermal properties of the clusters, the main  29 objects  in the {\tt \agn} simulations  were classified, at $z=0$, in CC and NCC systems. {In particular, those systems with a central entropy $K_0<60$ keV cm$^2$ and a pseudo-entropy $\sigma<0.55$ were considered as CCs, whereas those systems not fulfilling these conditions were instead classified as NCCs.}
In this way,  11 out of 29 halos are classified as CC clusters. The samples of CC and NCC clusters have been shown to be different in terms of  metallicity and thermodynamical profiles, showing a remarkably consistency with observations of these two cluster populations. 
In a similar way, based on the global dynamical state of the 29 main systems in the  {\tt \agn} simulations, we have classified them in 6 regular, 8 disturbed and 15 intermediate systems \citep[e.g.][]{Biffi_2016, Planelles_2017}. In order to do this classification we have considered a system to be regular when $\delta r < 0.07$ and $f_{sub} < 0.1$, where $\delta r$ and $f_{sub}$ represent, respectively, the centre shift between the cluster minimum potential and the cluster centre of mass, in units of its radius $R_{200}$, and  the mass fraction contributed by substructures within the same radius. Systems with larger values for both $\delta r $ and $f_{sub}$ are classified as disturbed, whereas those systems not satisfying concurrently  both criteria are labeled as intermediate cases. The samples of dynamically regular and disturbed systems have been also shown to differ in terms of the levels of hydrostatic equilibrium and gas clumpiness, especially in outer cluster regions \citep[see][for further details]{Biffi_2016,Planelles_2017}.

Note that the classification of clusters in dynamically regular or disturbed depends on the radius at which the dynamical state is defined \citep[e.g.][and references therein]{DeLuca2021}. Therefore, although the choice of this radius is quite arbitrary, it is important to keep this aspect in mind when discussing our results about the subsamples of dynamically regular and disturbed systems.

\subsection{Shock-finding algorithm}
\label{sec:shock_finder}

In general, the development of a shock in a cosmological simulation generates a discontinuity in all the hydrodynamical gas quantities (namely, gas density, temperature, pressure and entropy). The pre- and post-shock values of these quantities are connected to the strength of the shock through the Mach number, $\mathcal{M}=v_{s}/c_{s}$, where $v_{s}$ and $c_{s}$ are the shock speed  and  the sound speed ahead of the shock, respectively. 

Numerical methods developed to identify hydrodynamical shocks in cosmological simulations commonly rely on the Rankine-Hugoniot jump conditions \citep{Landau_1966}, which provide all the necessary information to calculate the shock Mach number. However, depending on the particular quantity and the particular way in which these conditions are applied, a number of shock-finding algorithms have been suggested. Thus, attending to the particular jump condition from which the shock Mach number is ultimately derived, there are methods based on the temperature, the density, the entropy or the velocity jump condition \citep[e.g.][]{Vazza_2009b}.  
Likewise,  depending on the way in which the direction of shock propagation is defined, there are two main numerical approaches:  the coordinate or directional-splitting methods, that follow the jump in a given hydrodynamical quantity along the coordinate axes of the computational domain \citep[e.g.][]{Ryu_2003, Vazza_2009, Planelles2013, Hong_2014, Hong_2015}, and the methods based on the local temperature or pressure gradients \citep[e.g.][]{Skillman_2008, Schaal_2015, Schaal_2016, Beck_2016}. Despite the obvious and unavoidable deviations between different shock-finding schemes and the distinct nature and properties of the cosmological simulations on which they are applied, all algorithms tend to produce similar properties of the shock waves in the LSS \citep[see, e.g.,][for a comparison study with different cosmological codes]{Vazza_2011b}.

In this work, in order to analyze the distribution, the evolution and the strength of shock waves in the simulations described in Section \ref{sec:simu}, we will employ the  shock-finding algorithm presented in \cite{Planelles2013} \citep[see also][for further details]{Sergio_2017}. {The definition of cells for these Lagrangian simulations will be described in Section \ref{sec:procedure}. Here we only outline the main procedure of the algorithm and defer the interested reader to the previous references for further details. }

The shock finder algorithm presented in \cite{Planelles2013} was developed to measure the strength of shocks in grid-based cosmological simulations. The code relies on a directional-splitting approach along the three coordinate axes and evaluates the shock Mach number from the temperature jump equation:
\begin{equation}
\frac{T_1}{T_2}=\frac{(5\mathcal{M}^2-1)(\mathcal{M}^2+3)}{16\mathcal{M}^2}\, ,
\label{eq:TJ}
\end{equation}
where $T_1$ and $T_2$ refer to the pre- and post-shock temperatures.

In our scheme, we first mark grid  cells as {\it tentative shocked}  if they fulfill the following conditions: 
\begin{equation}
(i)\, \nabla \cdot {\bf v} < 0\,; (ii)\, \nabla T \cdot \nabla S > 0\, , 
\label{eq:condi}
\end{equation}
where (i) ensures the  local gas velocity field ${\bf v}$ to be convergent and (ii) guarantees the same direction for the gradients of gas temperature $T$ and entropy $S$. 
Then, among all the tentative shocked cells,  the cell where $\nabla \cdot {\bf v}$ is minimum is identified as the first shock centre. Moving outwards from the shock centre, we define the extension of the shock along  each of the  three coordinate axes by looking for adjacent tentative shocked cells  where the pre- and post-shock temperature and density satisfy $T_2 > T_1 $\ and $\rho_2 > \rho_{1}$.
Once  the  furthest  shocked cells ahead (pre-shock) and behind  (post-shock) of the shock centre along  
each  coordinate axis are identified, the Mach number along each coordinate direction is computed substituting the corresponding temperatures $T_1$  and  $T_2$ in Eq.~\ref{eq:TJ}.
Following this approach shock discontinuities are spread, typically, over a few cells.
The Mach numbers obtained along the three coordinate directions are then combined to get the final strength of the first shocked cell:  $\mathcal{M}=(\mathcal{M}_{x}^{2}+\mathcal{M}_{y}^{2}+\mathcal{M}_{z}^{2})^{1/2}$, thus minimizing projection   effects   in  case   of   diagonal   shocks. 
After this step, this shocked cell is removed from the list of tentative shocked cells.
Then,  the whole process  is iteratively repeated focusing on the cell with the minimum value of $\nabla \cdot {\bf v}$.
  
With this procedure we can {identify} all  the  shocked  cells  within  the  computational  volume, obtaining  their Mach number and the typical shock surfaces associated to cosmological shock waves.
As an additional condition, in order to prevent noisy shock regions, it is a common practice to neglect all shocked cells with a Mach number lower than 1.3 \citep[e.g.][]{Ryu_2003, Planelles2013, Schaal_2015}.

The Rankine-Hugoniot jump conditions also provide an estimation of the amount of kinetic energy flux crossing a shock ($f_K$)  converted into thermal energy flux ($f_{th}$) in the post-shock region\footnote{Note that conversion of kinetic to thermal energy at shocks in SPH simulations is described by the prescription of artificial viscosity, which not necessarily coincides with the Rankine-Hugoniot jump conditions.}. The incoming kinetic energy flux at a shock is given by $f_K={\frac{1}{2}\rho_1(c_{s,1}\mathcal{M})^3}$, where $\rho_1$ and $c_{s,1}$ are the gas density and the sound speed in the pre-shock region, respectively. 
Written in this way, $f_K$ has units of energy flux, that is energy per unit time and per unit area. Therefore, the kinetic energy flux across shock surfaces could be computed as $F_K=f_K(\Delta x)^2$, being $\Delta x$ the spatial grid resolution.
Given the kinetic energy flux, the generated thermal energy flux can be estimated as follows:
\begin{equation}
f_{th}=\delta({\mathcal{M}})f_K\, ,
\label{eq:f_th}
\end{equation}
where $\delta({\mathcal{M}})$ is the gas thermalization efficiency at shocks. Following \cite{Kang_2007} for the case without a pre-existing CR component, $\delta({\mathcal{M}})$ can be computed as a function of the Mach number:
\begin{equation}
\delta({\mathcal{M}})=\frac{2}{\gamma(\gamma-1)\mathcal{M}^2 R}\left[\frac{2\gamma \mathcal{M}^2-(\gamma-1)}{(\gamma+1)}-R^{\gamma}\right]\, ,
\label{eq:delta_m}
\end{equation}
where $\gamma$ is the adiabatic exponent and $R$ depends on the  jump in density:
\begin{equation}
R\equiv\frac{\rho_2}{\rho_1}=\frac{\gamma+1}{\gamma-1+2/\mathcal{M}^2} \, .
\label{eq:R}
\end{equation}

Besides gas thermalization, part of the energy flux at a shock can also be transferred into cosmic rays. Indeed, by means of the DSA mechanism, cosmic rays can also be accelerated and injected at shocks. The efficiency of this injection depends on the shock Mach number, on the degree of Alfv\'en turbulence and on the direction of the magnetic field relative to the shock surface. \cite{Kang_2007} inferred limits for the efficiency of CR acceleration, $\eta({\mathcal{M}})$, providing fitting functions for the cases with and without a pre-existing CR component. In our analysis, once shocks are identified we will use these fitting formulae, for the case without a pre-existing CR component, to estimate the flux of energy transferred into CRs:
\begin{equation}
f_{CR}=\eta({\mathcal{M}})f_K\, .
\label{eq:f_th}
\end{equation}

Note that the treatment of energy dissipation at shocks by means of an approximated post-processing approach, like the one described above, presents a number of simplifications and caveats. However, it will allow us to get some upper limits for the efficiency of energy dissipation at shocks.   
It is also important to keep in mind that the Rankine-Hugoniot jump conditions provide a description of the strength of shock waves in the case of ideal shocks. Since shocks in simulations are far from being ideal, any numerical procedure based on these equations will suffer from inevitable uncertainties.

\subsection{Numerical procedure}
\label{sec:procedure}

Given the characteristics of our shock-finder algorithm, in order to obtain the distribution of shock waves and their associated Mach numbers, the code needs to be applied onto a computational grid where the gas main quantities are stored in cells of any given resolution. 
Therefore, before running the shock-finder we need to convert all the information provided by our simulations from the distribution of particles to a proper computational grid. In order to compute the Mach number within different radial apertures from the cluster centre we proceed in the following way.  For each selected simulated cluster, with virial radius $R_{vir}$,
we build a cubic box, with centre coinciding with the cluster centre, and with a side length of $8\times R_{vir}$. The box is then discretized with a given number of cubical cells ($N=128^3, 256^3$, 5$12^3$ cells). 
On average, for the whole sample of 29 clusters  in the {\tt \agn} run at $z=0$, when we consider a number of grid cells equal to $N=128^3, 256^3$, $512^3$ we obtain, respectively,  a  mean spatial resolution of $\Delta x\simeq 144, 72$ and $36\, \kpch$.
Then, we select all the gas particles within a distance of $4\times R_{vir}$ from the halo centre and we interpolate the main gas properties onto the regular grid \citep[see][for a recent comparison of different smoothing methods]{Rottgers_2018} by spreading them accordingly to the same Wendland C4 kernel used for the SPH computations in the simulations \citep[see][]{Beck_2015}.
After this, we apply the shock finder in order to compute the shock Mach number for several radial apertures from the cluster centre. In particular, for each halo we have considered 4 different radial apertures: $R_{core}\equiv 0.05\times R_{180}, R_{2500}, R_{500}$ and $R_{vir}$, {although throughout this work we will mainly focus on $R_{500}$ and $R_{vir}$}. Once this procedure is applied to the main 29 clusters, we analyze the distribution of shock Mach numbers as a function of redshift, cluster mass, cluster cool-coreness, cluster dynamical state, radial aperture, spatial grid resolution {and physics included in the simulations.}

Unless otherwise specified, in the following we will employ as our reference results those obtained for the {\tt \agn} simulations when smoothing each computational domain onto a regular grid with {$128^3$ cells}. {In Appendix \ref{app:resol} we explore the effects of grid resolution on our results  and we justify our choice}. The selected spatial grid resolution is in line with the resolution employed in some recent works on the topic \citep[e.g.][]{Vazza_2010, Vazza_2011b, Vazza_2013}.

\section{Results}
\label{sec:results}

In Sections \ref{sec:shock_distribution} and \ref{sec:scaling_relations} we present a global analysis of different shock statistics as obtained for the whole sample of 29 regions in the {\tt \agn} simulation at our reference grid resolution. We also analyze the dependence of the different quantities on cluster mass and redshift.
In contrast, in Appendix \ref{app:clusters_and_shocks} we focus our analysis on a selected subsample of 6 simulated clusters and we put special emphasis on the connection between shock waves and the cluster cool-coreness and dynamical state. 

\subsection{Global shock cell distribution}
\label{sec:shock_distribution}

\begin{figure*}
\begin{center}
{\includegraphics[width=16cm]{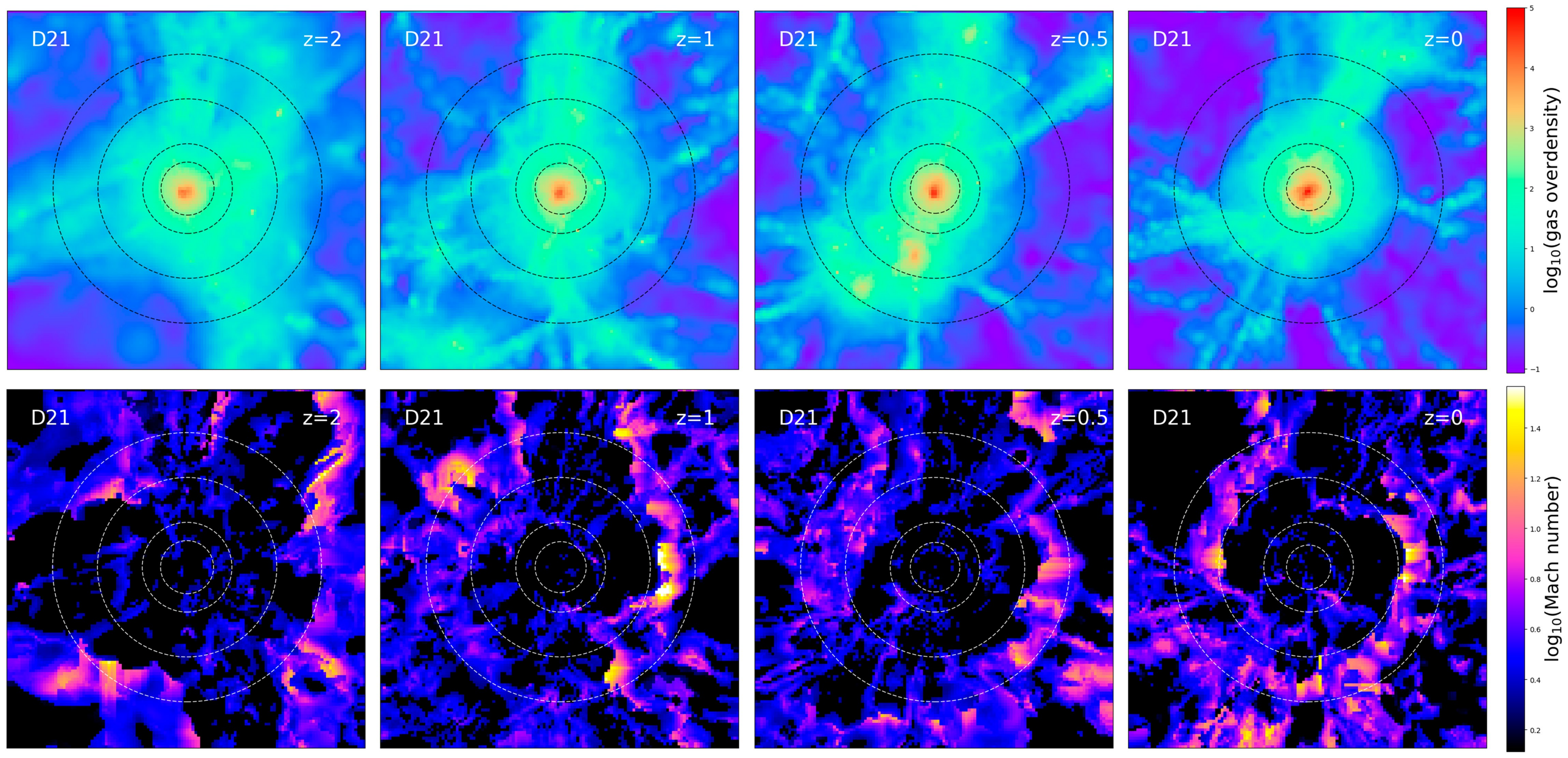}}
\caption{2-D projections of the gas overdensity (upper panels) and the Mach number distribution (lower panels) around one of the most massive clusters in our sample at different redshifts. Panels from left to right show the projections at $z=2, 1, 0.5$ and 0, respectively. Each map, projected along the z-axis, has a length of $8\times R_{vir}$ at each redshift and a projection depth of 7 cells centred on the cluster position. Circles on the maps represent, respectively, $R_{500}$, $R_{vir}$, $2\times R_{vir}$ and $3\times R_{vir}$ of the central cluster.}
\label{fig:z_evol_maps}
\end{center}
\end{figure*}

We start by analysing the global shock cell distribution around D21, one of the most massive clusters in our {\tt \agn} simulation. Figure \ref{fig:z_evol_maps} shows 2-D projections along the z-axis of the gas overdensity (upper panels) and the Mach number distribution (lower panels) at different redshifts (panels from left to right show the projections at $z=2, 1, 0.5$ and 0, respectively). Each map has an extent of $8\times R_{vir}$ at each redshift and a projection depth of 7 cells centred on the cluster position. Cluster D21, with $M_{vir}=14.26\times10^{14}\msunh$ and $R_{vir}=2.37\mpch$, has been classified as dynamically relaxed and it is a CC cluster. As can be seen in the density maps, the cluster is already well identified at $z=2$. The mean Mach number within the whole region is around $\sim4$ throughout the redshift evolution (see also Fig.~\ref{fig:z_evol}). However,  as evolution proceeds 
the Mach number of shocks around the central cluster increases, developing a significant high-Mach number accretion shock  around the cluster centre at $z=0$. At the same time, as redshift decreases, the distribution of shocks becomes somewhat more volume-filling (see Fig.~\ref{fig:z_evol}). Moreover, according to these maps, high-Mach number external shocks seem to be clearly dominant, in terms of strength and extension, over low-Mach number internal ones. 
To understand this behaviour it is interesting to note that accretion shocks from voids and  low-density regions onto the cosmic web are very volume-filling. In addition, as evolution proceeds void regions expand and increase their size, making the distribution of low-Mach number shocks within them to appear somewhat more diluted. Concurrently, the temperature of the gas in voids decreases, producing a significant difference between the temperature in voids and that in collapsed regions and, therefore, generating stronger shocks around collapsed structures.

\begin{figure*}
\begin{center}
{\includegraphics[width=15cm]{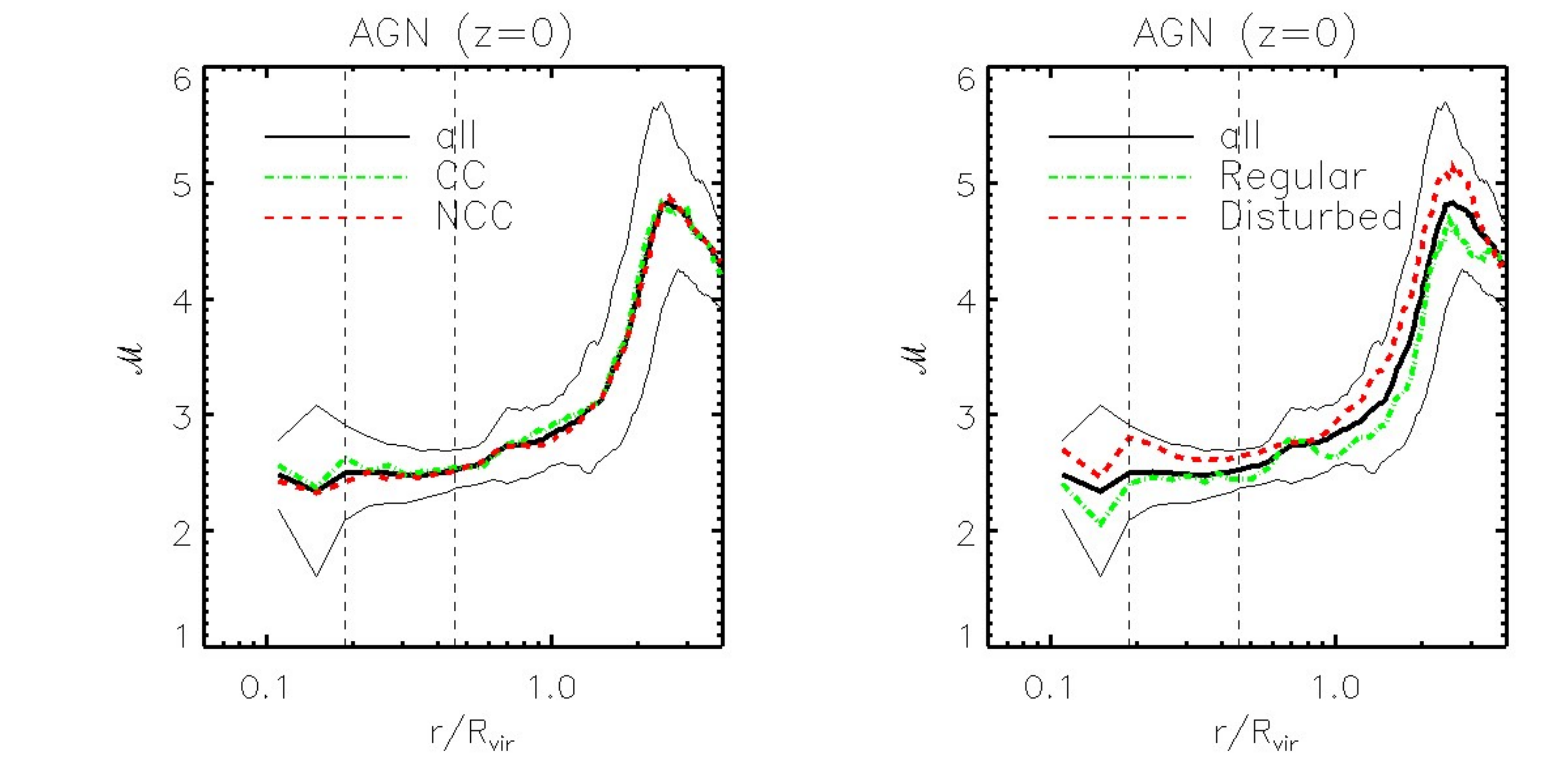}}
\caption{Volume-averaged radial profiles  of the mean Mach number within $4\times R_{vir}$ for the whole sample of 29 central clusters in the {\tt \agn} simulation at $z=0$ (thick black continuous lines in both panels). Results for the subsamples of CC/NCC clusters and for the regular/disturbed clusters are shown, respectively, in the left and right panels. Thin black continuous lines in both panels represent 1-$\sigma$ standard deviation around the mean global profile. Dashed vertical lines stand for the mean values of $R_{2500}$ and $R_{500}$ as obtained for the whole sample of clusters.}
\label{fig:radial_prof_state}
\end{center}
\end{figure*}

In Appendix \ref{app:physics} we compare the distribution at $z=0$ shown in Fig.~\ref{fig:z_evol_maps} for cluster D21 with the same distribution as obtained in the {\tt \nr} and {\tt \csf} simulations (see Fig.~\ref{fig:d21_physics_big_region}). On average, the shock configurations are similar with comparable Mach numbers and locations, confirming the fact that the high-Mach number external shocks shown in  Fig.~\ref{fig:z_evol_maps} are indeed accretion shocks, with the exception  of small differences in the Mach number distribution between the  {\tt \nr} and the {\tt \agn} simulations, as discussed more in Appendix B.  
As inferred from previous simulations \citep[e.g.][]{Schaal_2015,Schaal_2016}, outer accretion shocks show typically quasi-spherical shapes and are  found at distances of $\sim1.5-2\times R_{vir}$. According to Figs.~\ref{fig:z_evol_maps} and \ref{fig:d21_physics_big_region} (see also Fig.~\ref{fig:radial_prof_state}), the accretion shock developed at $z=0$ around cluster $D21$ also shows a quasi-spherical pattern, it wraps the LSS around the cluster and it is   located at $\sim 2.5\times R_{vir}$ from the cluster centre.

\begin{figure*}
\begin{center}
{\includegraphics[width=7.5cm]{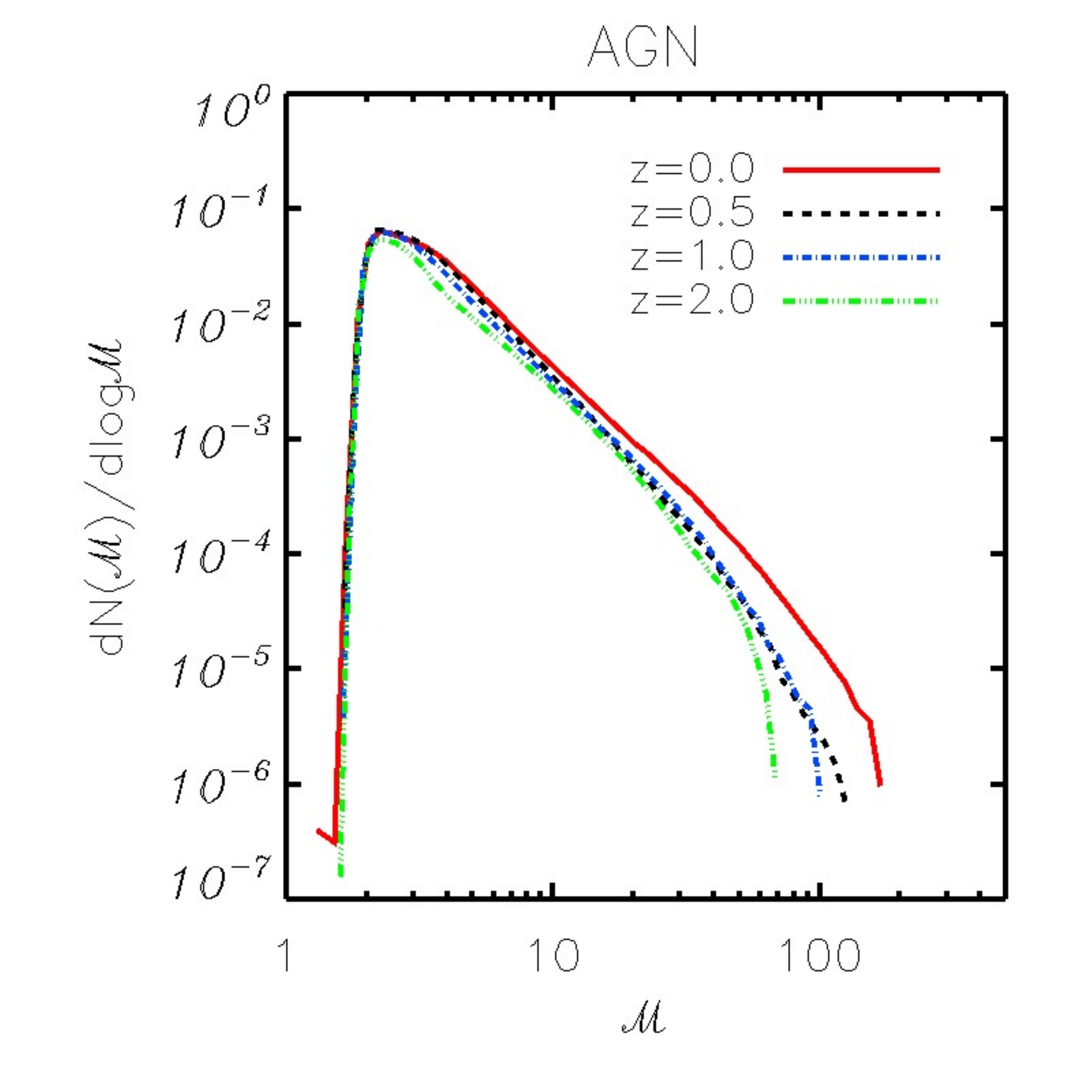}}
{\includegraphics[width=7.5cm]{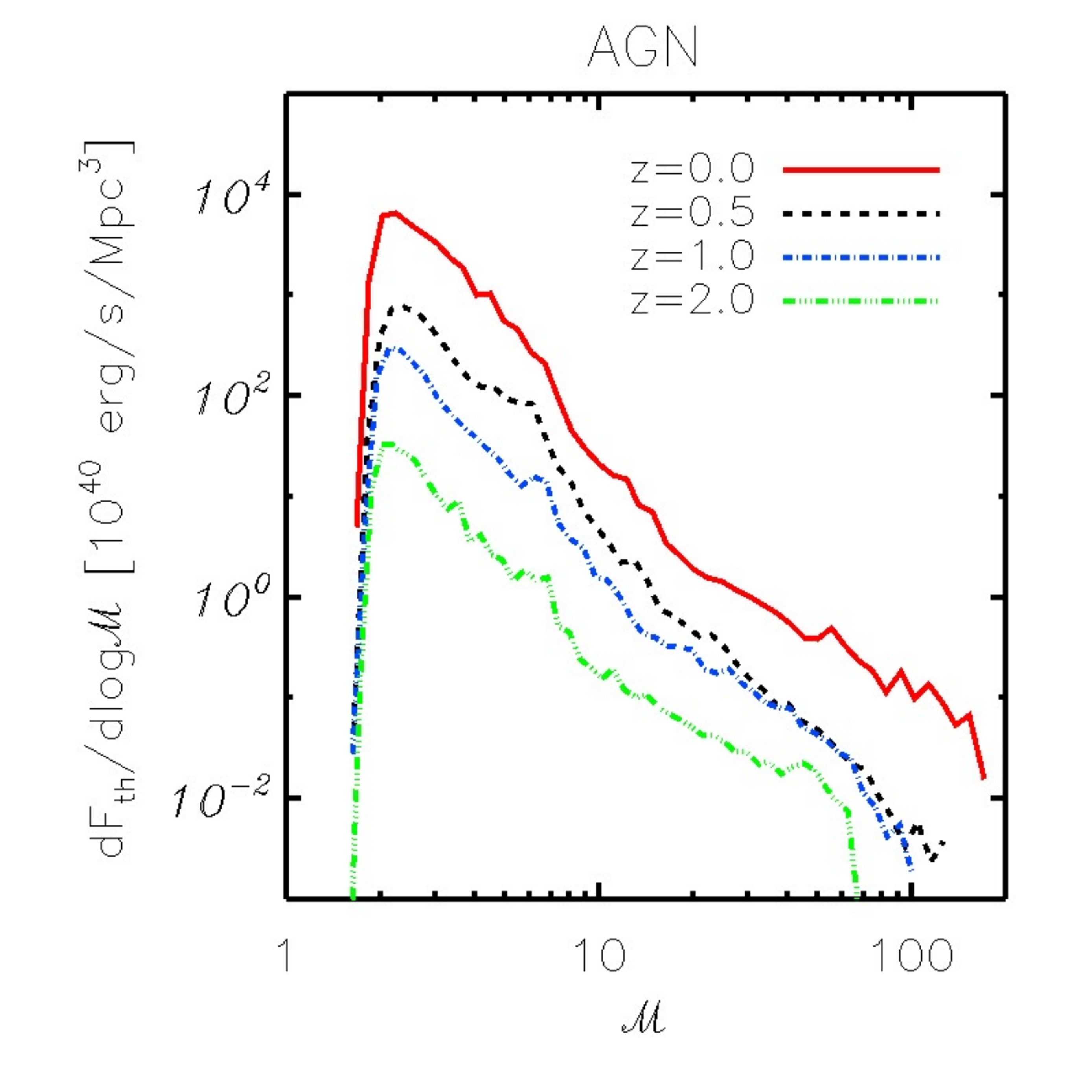}}
\caption{{\it Left panel:} Mean shock cell distribution function obtained for the sample of 29 clusters in the  {\tt \agn} simulation as a function of  redshift. {\it Right panel:} Distribution of the thermal energy flux through shocks (per unit volume) for the sample of 29 clusters in the  {\tt \agn} simulation as a function of  redshift.}
\label{fig:distri_all_halos}
\end{center}
\end{figure*}

To have an idea of the way in which shocked cells of different strengths are distributed through the computational volume, in the left panel of Fig.~\ref{fig:distri_all_halos} we show the mean shock cell distribution function (SDF, from now on) obtained for the whole sample of 29 clusters in the  {\tt \agn} simulation at different redshifts. 
From the inspection of this figure at $z=0$ (red continuous line), we can extract several broad conclusions. As shown in previous studies \citep[e.g.][]{Skillman_2008, Vazza_2009,Planelles2013}, on average most of the computational volume is filled with low-Mach number shocks ($\mathcal{M}\mincir 3$, blue-violet colors in Fig.~\ref{fig:z_evol_maps}), which tend to be located within the virial radius. In contrast, stronger shocks with higher Mach numbers (orange-yellow colors  in Fig.~\ref{fig:z_evol_maps}) are less numerous and tend to be found in outer cluster regions (see also Fig.~\ref{fig:radial_prof_state}). According to this behaviour, the SDF shows a peak at around $\mathcal{M}\sim2$ and decreases beyond that value. As already discussed by some other authors \citep[e.g.][]{Ryu_2003}, internal shocks within clusters result from a number of processes such as substructure mergers, accretion of warm-hot gas along filaments, or internal turbulent flow motions. On the contrary, external shocks around clusters are associated to large-scale accretion phenomena, especially when the cold gas from lower-density regions accretes onto the hot ICM. The contrast, in terms of density and temperature, between shocked and unshocked regions within the different environments where internal and external shocks are developed makes them to have very different strengths.    

As  for the dependence of the SDF on redshift, we show that the general shape of the mean SDF remains quite stable already from $z=2$. However, the amount of shocked cells within the computational domain slightly increases with decreasing redshift, reaching a mean fraction of shocked cells of $\sim13$ per cent at $z=0$. In this sense, as the redshift decreases the number of shocked cells within the considered regions augments, giving raise, however, to a mild increase in the amount of weak internal shocks. In a similar way, as the  cosmic evolution proceeds, the temperature of the IGM, especially in low-density regions, decreases, explaining the increasing high-Mach number end of the SDF \citep[e.g.][]{Vazza_2009,Schaal_2015,Schaal_2016}.

For completeness, the right panel of Fig.~\ref{fig:distri_all_halos} shows the mean z-evolution of the distribution of the thermal energy flux through shocks for the sample of 29 clusters in the  {\tt \agn} simulation. Overall, at any redshift, the maximum of the gas thermalization by shocks is found at $\mathcal{M}\sim2$, consistent with previous studies \citep[e.g.][]{Ryu_2003,Pfrommer_2006, Skillman_2008, Vazza_2009}. In addition, independently of the considered redshift, the shape of the thermal energy distribution is quite alike, showing a steep negative dependence with the Mach number. In broad agreement with previous results \citep[e.g.][]{Pfrommer_2007, Vazza_2009}, we find the slope of this distribution to be within $\alpha\simeq -4$ ($z=0$) and $\alpha\simeq -3$ ($z=2$). Despite the high Mach number of external shocks, given the low-density environment where they are usually developed, the associated kinetic energy flux is very small, making them energetically less important. On the contrary, low-Mach number shocks, mainly developed within collapsed and denser structures, have larger values of $f_{K}$ and, therefore, they are more relevant in thermalizing the ICM and producing CRs. As a consequence, most of the  total thermal energy flux is processed by shocks with relatively low Mach numbers \citep[$\mathcal{M}\mincir10$; e.g.][]{Vazza_2011b}.

In order to explore the radial distribution of shocks around the main clusters of each simulated region, Fig.~\ref{fig:radial_prof_state} shows, at $z=0$, the averaged radial profiles of the mean Mach number  for the whole sample of 29 objects in the {\tt \agn} simulation  (black continuous line). The mean radial profiles corresponding to the subsamples of CC/NCC  and regular/disturbed systems are also shown for comparison (left and right panels, respectively). 

In general, when we analyze the mean global profile, we find that the mean Mach number radial distribution is quite flat out to $\sim0.8\times R_{vir}$, showing values within the range $\sim 2.2-2.5$. At radial distances  above  $\sim0.8\times R_{vir}$ the profiles raise significantly, reaching values as high as $\mathcal{M}\sim5$ in the outer cluster regions.
This trend is in broad agreement with results obtained in previous studies \citep[e.g.][]{Vazza_2009b,Vazza_2010,Vazza_2011b}. In particular, \citet{Vazza_2011b} performed a comparison of the radial Mach number distribution in two massive clusters as obtained with two grid-based cosmological codes and with the SPH code GADGET-3. Comparing with our results, they also found a mean Mach number value below $\sim3$ inside the virial radius \citep[see also, e.g.][]{Ha_2018} and a very similar radial shock cell distribution, at least within $\sim0.5 R_{vir}$. However, {beyond} the virial radius, they showed  that grid codes tend to identify thinner and stronger shocks than  GADGET-3, producing a much sharper increase of the profiles. This abrupt rise in the strength of shocks clearly marks the transition between the outskirts of clusters  and the low-density IGM.
The particular shape of the profiles in these outer cluster regions  could be affected by a number of factors such as the  simulation numerical scheme, the specific shock-detecting approach, the different cluster environment derived from the intrinsic difference between cosmological and zoom-in simulations, or the grid resolution and/or the gas sampling procedure (especially in outer and underdense cluster regions where, by construction, SPH is not very good at sampling).
For all these reasons, a precise comparison  between different simulations is not straightforward. Nevertheless, even with these weaknesses, results are in confortable agreement with those obtained using grid codes.

As for the connection between the distribution and strength of shocks and the cluster cool-coreness and global dynamical state, some previous studies attempted to analyze the relation between relaxed and unrelaxed (merging) clusters with the strength of their associated shock waves 
\citep[e.g.][]{Vazza_2011, Vazza_2012}. Overall, these studies found similar shock cell distributions from cluster to cluster with only small deviations connected to their particular dynamical state, especially above the virial radius.

In our case, according to the results shown in the left panel of Fig.~\ref{fig:radial_prof_state}, we do not obtain any difference between the mean Mach radial profiles when we look separately to the samples of CC and NCC clusters. This result is not surprising, since the classification of clusters according to their cool-coreness is based on a much smaller radial aperture \citep[$R_{core}\equiv 0.05\times R_{180}$;][]{Rasia_2015}, which is completely smoothed out at the considered resolution. As discussed in Appendix \ref{app:resol}, given the numerical approach we follow to smooth the particles' distribution onto a computational grid, it would be risky to artificially increase the spatial grid resolution  to try to  `resolve' the core of clusters. 

When we analyze instead the mean radial profiles obtained  for the subsamples of relaxed and disturbed systems (right panel of Fig.~\ref{fig:radial_prof_state}) we obtain a  larger difference between them: disturbed clusters clearly show a higher mean Mach number than regular ones throughout the radial range. A similar behaviour was also obtained for our sample of regular and disturbed systems when we compared their degree of deviation from the hydrostatic equilibrium condition or their levels of gas clumpiness  \citep[see][respectively]{Biffi_2016, Planelles_2017}. The trend shown in the right panel of Fig.~\ref{fig:radial_prof_state}, which was somehow expected in our sample, is also in broad agreement with previous analyses \citep[e.g.][]{Vazza_2009b}.

\subsection{Shock waves scaling relations}
\label{sec:scaling_relations}

\begin{figure*}
\begin{center}
{\includegraphics[width=7.5cm]{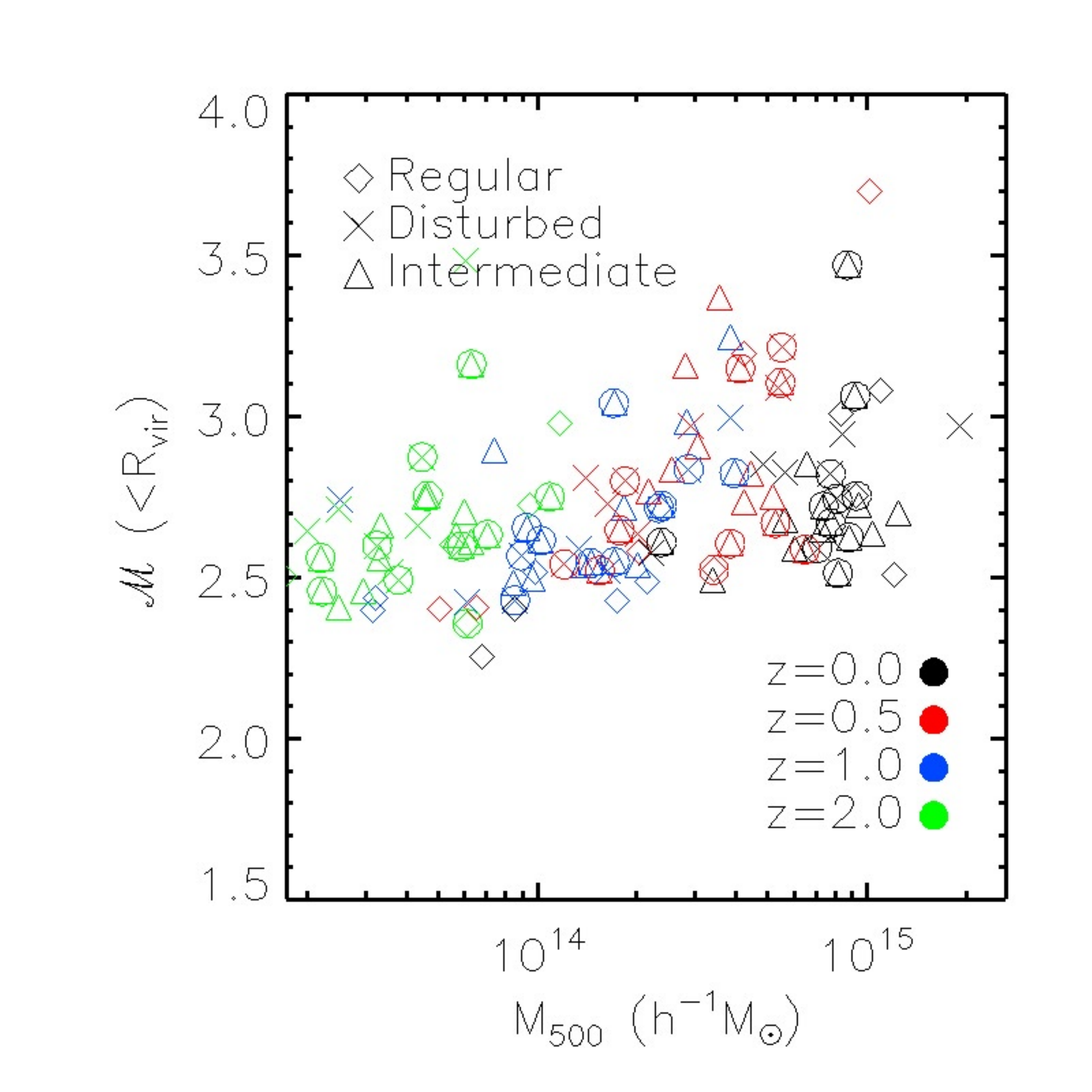}}
\hspace{0.5cm}
{\includegraphics[width=7.5cm]{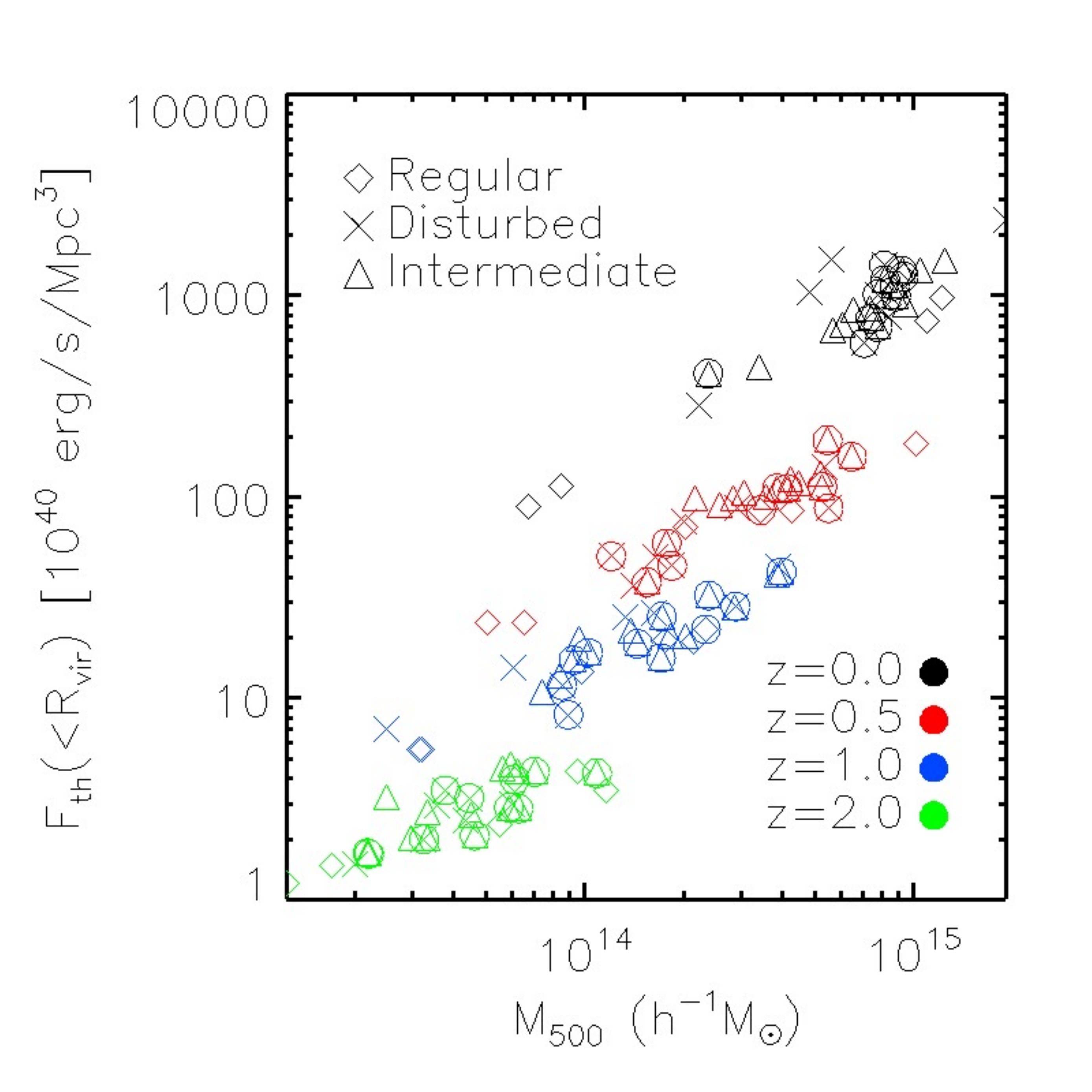}}
\caption{{\it Left panel:} Mean volume-weighted Mach number within $R_{vir}$ as a function of the cluster mass $M_{500}$ for the whole sample of 29 central clusters in the {\tt \agn} simulation at redshift $z=0, 0.5, 1$ and 2. The value of $\mathcal{M}(<R_{vir})$ is shown for the populations of regular/disturbed/intermediate systems as specified in the legend. According to the core cluster properties, CC clusters are marked with an additional circle around the symbols shown in the legend. {\it Right panel:} Same results as in the left panel but for the mean thermal energy (divided by the volume of clusters), $F_{th}$, within $R_{vir}$.}
\label{fig:corr_all_halos}
\end{center}
\end{figure*}

\begin{table}
\begin{center}
  \begin{tabular}{lcccc}
  \hline
  Sample & $\mathcal{M}_{vir}$ & $F_{th, vir}$  & $F_{CR, vir}$ \\
               &                                  &  $[10^{45} erg/s]$ &  $[10^{45} erg/s]$ \\
  \hline
  All                & 2.74  &  1.65  &  0.74 \\
  CC              & 2.78  &  2.77   &  1.25 \\
  NCC           & 2.71  & 0.96    & 0.43 \\
  Regular      & 2.67  & 1.15   & 0.44 \\
  Disturbed   & 2.79  & 2.28  & 1.13 \\
\hline
\end{tabular}
\caption{Mean values at  $z=0$ of the Mach number ($\mathcal{M}$), 
the thermal and the CR energy flux at shocks ($F_{th}$ and $F_{CR}$, respectively) within $R_{vir}$ for the whole sample of clusters in the {\tt \agn} simulation and for the subsamples of CC/NCC and regular/disturbed systems.}
\label{tab:mean_val}
\end{center}
\end{table}

Galaxy cluster scaling relations are extremely useful to estimate cluster masses from cluster observables. In {the left panel of} Fig.~\ref{fig:corr_all_halos} we show the mean Mach number within  $R_{vir}$ as a function of the cluster mass $M_{500}$ for the whole sample of 29 central clusters in the {\tt \agn} simulation at redshifts $z=0, 0.5, 1$ and 2.  
The value of $\mathcal{M}(<R_{vir})$ is shown for the populations of dynamically regular, disturbed and intermediate systems as specified in the legend. Moreover, CC clusters are marked with an additional circle around the main symbols, being all the clusters without this circle NCC clusters.
Similarly,  in the right panel of Fig.~\ref{fig:corr_all_halos}  the mean thermal energy flux, $F_{th}$,  within $R_{vir}$  is shown. Table \ref{tab:mean_val} shows the mean values of these quantities at $z=0$ for the different cluster samples. For the sake of completeness, the mean energy flux inverted in CR acceleration is also given. 

On average, when all redshifts are considered, the mean Mach number value within the virial radius is almost flat as a function of cluster mass, with a  significant dispersion for all redshifts ($\mathcal{M}(<R_{vir})\sim[2.2, 3.8])$.  However, if we consider the whole cluster sample at $z=0$, the Pearson's correlation coefficient between the Mach number values within the virial radius and the mass $M_{500}$ is $\sim0.47$, indicating a moderate correlation. Instead, when we consider the subsamples of regular and disturbed systems, we obtain a positive correlation with values of the Pearson's correlation coefficient above $\sim0.7$. We have checked that this trend holds from $R_{200}$ and beyond. However, given that the considered mass range is quite narrow and the cluster sample is relatively limited, we should be cautious when deriving conclusions on this regard.
As suggested by  the profiles shown in Fig.~\ref{fig:radial_prof_state}, mild differences between dynamically regular and disturbed systems are observed throughout the radial range, contributing to a difference between the mean $\mathcal{M}(<R_{vir})$ of regular and disturbed systems  which is not found between that of CC and NCC clusters (see also Table \ref{tab:mean_val}).

\begin{figure*}
\begin{center}
{\includegraphics[width=7.5cm]{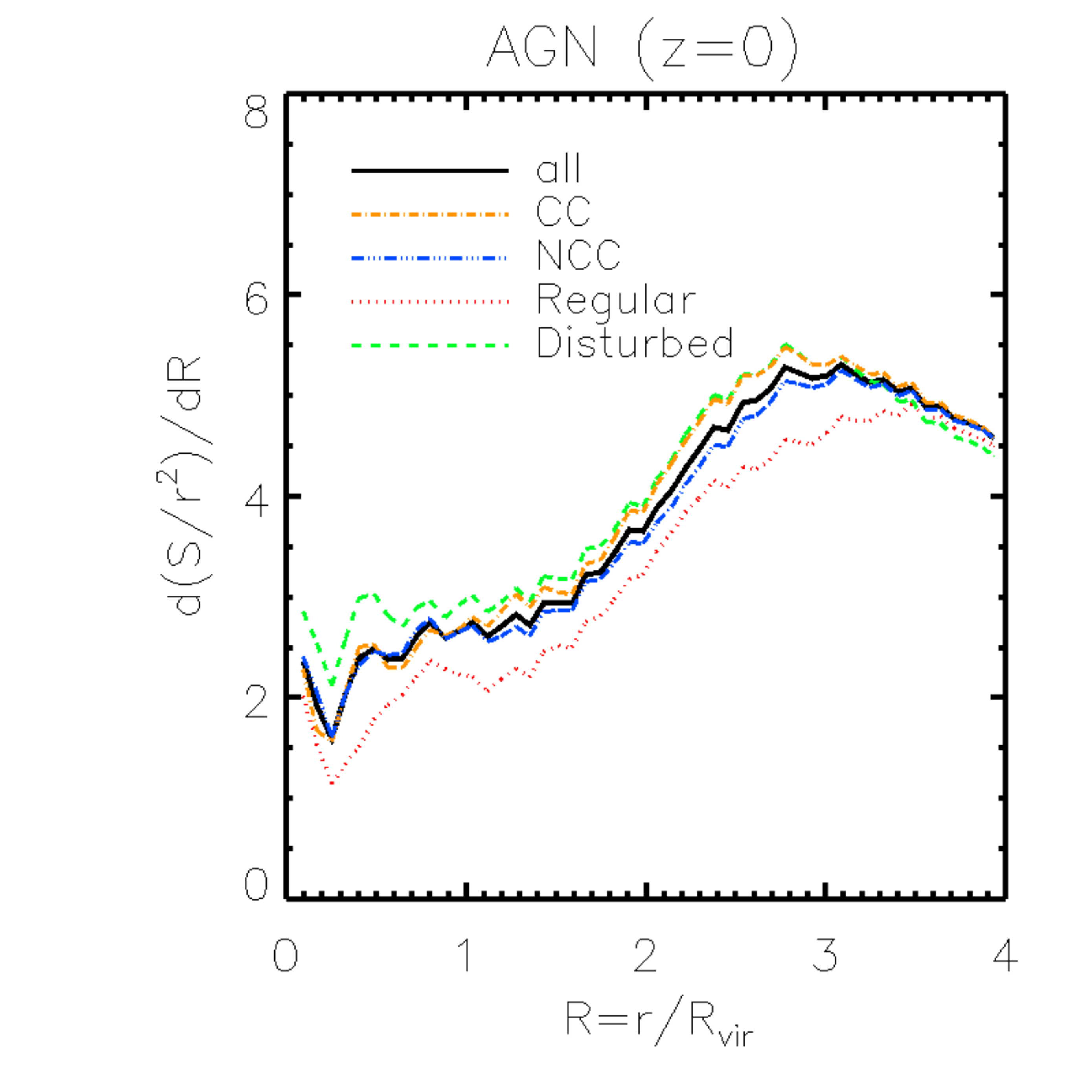}}
{\includegraphics[width=7.5cm]{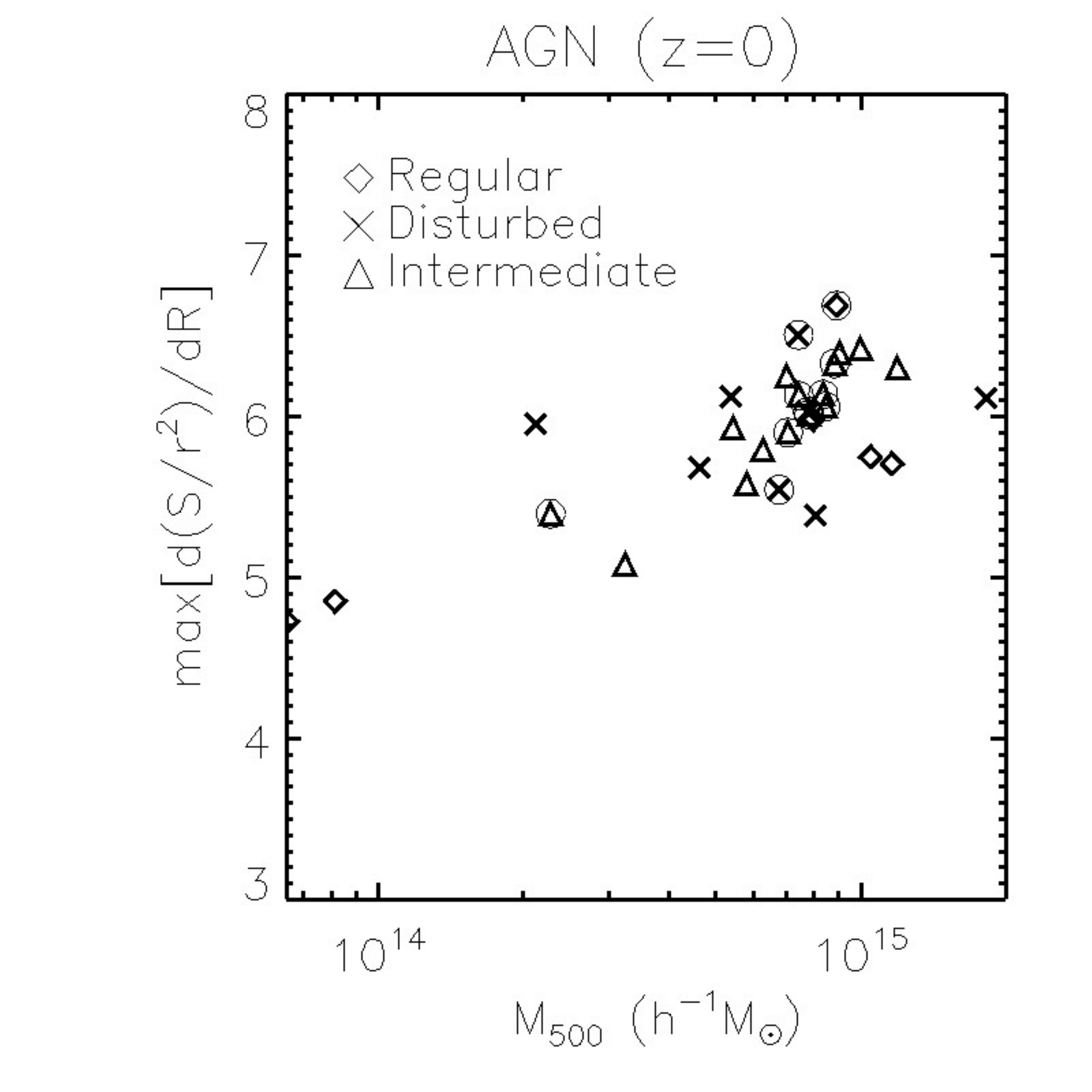}}
\caption{{\it Left panel}: Mean radial shock surface distribution, out to $4\times R_{vir}$, for the whole sample of 29 clusters in the {\tt \agn} simulation at $z=0$ (black continuous line), and for the subsamples of relaxed/disturbed (red and green lines, respectively) and CC/NCC clusters (orange and blue lines). {\it Right panel}: Maximum value of the radial shock surface distribution, $max(d(S/r^2)/dR$, as a function of $M_{500}$ as obtained for the whole sample of clusters in the  {\tt \agn} simulation at $z=0$. Regular, disturbed and intermediate systems are marked according to the symbols shown in the legend, whereas CC clusters are marked with an additional circle around those symbols.}
\label{fig:radial_prof_all}
\end{center}
\end{figure*}

On the contrary, as shown in the right panel of Fig.~\ref{fig:corr_all_halos}, the mean thermal energy flux within $R_{vir}$ shows, at fixed redshift, some correlation with cluster mass although somewhat scattered. It is only combining different redshifts that a stronger correlation shows up. However,  this seems to be due to the redshift dependence. Indeed, as discussed by other authors \citep[e.g.][]{Ryu_2003}, massive galaxy clusters are supposed to represent the main sites for energy dissipation at shocks. Therefore, since the total energy processed by shocks in a simulated volume depends on the number of clusters, we should be cautious when comparing these results with preceding works \citep[e.g.][]{Ryu_2003,Pfrommer_2006,Vazza_2009b}.  
According to the results shown in Table \ref{tab:mean_val}, at $z=0$ our sample of CC clusters have larger mean values of $F_{th, vir}$ than the NCC systems. In a similar way, as for the global dynamical state of clusters, disturbed objects show larger values of the mean thermal energy flux than regular ones. Since all these cluster subsamples have similar mean virial Mach numbers, differences could be also connected with a different amount of gas density contributing to the dissipation of energy.

Therefore, according to the global analysis of Fig.~\ref{fig:corr_all_halos}, our results suggest that, provided that we have a significant cluster statistics, these relations could be used as potential cluster mass proxies.

A complementary way to analyze the radial distribution of shocks around clusters is given in the left panel of Fig.~\ref{fig:radial_prof_all}, that shows the radial shock surface distribution, out to $4\times R_{vir}$, for the whole sample of clusters in the {\tt \agn} run at $z=0$ (black continuous line). 
This quantity provides an approximated estimation of the surface of shock cells, $S$, at their locations, $r$.
For the sake of comparison, we also report the mean shock surface distribution obtained for the subsamples of CC/NCC and regular/disturbed systems. As shown in previous studies \citep{Schaal_2015, Schaal_2016}, this distribution provides information on the location of the main shock surfaces within and around the cluster virial region and it is, therefore, very useful to estimate the position and shape of external accretion shocks. Within the virial radius, the different mean shock surface distributions show a number of peaks or bumps which make the profiles quite irregular. These internal peaks could be connected with a number of internal shock events, such as shocks developed as a consequence of merging substructures, flow motions or feedback processes. However, above $R_{vir}$ all mean profiles tend to smoothly increase out to $\sim2.5-3\times R_{vir}$, where the curves show their maximum value. The position of this maximum could be connected with the radius (provided that accretion shocks are assumed to be spherical) of the main external accretion shock around the central clusters. It is interesting to note that there is a clear distinction between the mean  shock surface distribution associated to regular and disturbed systems, with the latter showing the highest values throughout the radial range. 
On the contrary, we have not found such a strong dependence of $d(S/r^2)/dR$ on the clusters' core properties.
Interestingly, we have obtained a moderate correlation (with a Pearson's correlation coefficient of $0.6$) between the maximum value of the distribution and the mass of the associated systems. 
At $z=0$, by fitting linearly the relation max$(d(S/r^2)/dR)=f(log(M_{500}))$ (see right panel of Fig.~\ref{fig:radial_prof_all}), we have obtained for the whole sample of clusters {in the  {\tt \agn} run} a slope {$\alpha=1.1\pm0.2$}, although with some scatter\footnote{{Similarly, we have obtained for the reduced sample of clusters in the {\tt \csf} and {\tt \nr} simulations a value of $\alpha=1.1\pm0.9$ and $\alpha=1.3\pm0.9$, respectively.}}. This trend points towards a connection between the extension (or surface filling factor) of external accretion shocks developed during the collapse of  structures and the mass of the final formed systems. Assuming that future observational radio facilities, such as the {\it Square-Kilometre Array}\footnote{https://www.skatelescope.org/}  \citep[\ska; e.g.][]{Acosta-Pulido2015}, will be able to routinely detect large-scale shock waves, correlations like the one presented here, {which is rather independent on the physics included,} could become a complementary tool to estimate cluster masses \citep[e.g.][]{Planelles2013}.

\section{Summary and Conclusions}
\label{sec:conclusions}

{The formation and evolution of cosmic shock waves, inherent to cosmic structure formation and evolution processes,  play a main role in the energy distribution and thermalization of the IGM.
In this work we have analyzed the distribution of these shocks in the Dianoga set of simulated galaxy clusters obtained from a set of zoom-in simulations performed with the SPH code GADGET-3. These simulations account for the effects of radiative cooling, star formation, SNe and AGN feedback. Previous analyses of these simulations have shown an excellent agreement with a number of observed properties of the intra-cluster medium, such as the X-ray and SZ scaling relations, the metal distribution, or the thermal and dynamical properties of CC and NCC clusters \citep{Rasia_2015, Biffi_2017, Planelles_2017, Biffi_2018, Truong_2018,Truong_2019}. We have performed a further analysis of these simulations to explore in detail the distribution and evolution of the strength of shocks as a function of cluster mass, redshift and feedback processes, paying special attention to the connection between the shock cell distribution and the cool-coreness (CC/NCC) and global dynamical state (regular/disturbed) of clusters. To perform this analysis, a grid-based shock-finding algorithm has been applied in post-processing to the outcomes of the simulations. In the following we summarize our main findings.

\begin{itemize}
\item In general, as cluster evolution proceeds, a significant high-Mach number external accretion shock is developed around simulated clusters at $z=0$. These outer accretion shocks show typically quasi-spherical shapes and are  found at distances of $\sim(2-2.5)\times R_{vir}$ from the cluster centre.

\item As already shown in previous works, most of the computational volume is filled with low-Mach number shocks ($\mathcal{M}\mincir 10$), which tend to be located within the cluster boundaries, while stronger shocks with higher Mach numbers are less numerous and tend to be found in outer cluster regions.

\item Low-Mach number shocks, mainly developed within collapsed and dense environments, are more relevant in thermalizing the ICM and producing CRs. As a consequence, most of the  total thermal energy flux is processed by relatively low-Mach number shocks  \citep[$\mathcal{M}<10$; e.g.][]{Vazza_2011b}, while stronger external shocks are energetically less important.

\item While the mean Mach number radial distribution within a cluster is quite flat out to $\sim0.8\times R_{vir}$ (with mean values within $\sim2.2-2.5$), the profiles raise significantly in outer cluster regions, reaching values as high as $\mathcal{M}\sim 5$.

\item We do not find any relevant difference between CC and NCC clusters in terms of the shock Mach number radial distribution. However, according to the clusters' dynamical state, disturbed systems tend to show stronger shocks than regular ones throughout the clusters' volume.

\item We find that, in general, massive clusters tend to show higher mean Mach numbers within the virial radius than less massive ones and, moreover, their associated shock waves are more efficient thermalizing the IGM. This trend is even stronger when we consider the subsamples of regular and disturbed systems.

\item From the analysis of the shock surface distribution function we have obtained a moderate correlation between the extension  of external accretion shocks developed during the collapse of  structures and the mass of the final formed systems (see also Appendix \ref{app:clusters_and_shocks}). We expect this  correlation, which is independent of the physics included in the simulations, to become in principle a way to infer cluster masses.

\item As for the redshift evolution of the shock cell distribution we find that the fraction of cells hosting shocks in the different simulated volumes slightly increases with decreasing redshift (from $\sim46$ per cent at $z=2$ out to $55-60$ per cent at $z=0$; see Appendix \ref{app:clusters_and_shocks}). On the other hand, the mean Mach number remains quite flat  for all clusters ($3\mincir\mathcal{M}\mincir 4$). 

\item As a consequence of the different radiative and feedback processes included in the {\tt \nr}, {\tt \csf} and {\tt \agn} simulations, some differences in the distribution of shocks within the clusters' virial region are clearly detectable. On the contrary, the large-scale Mach number distribution, which is mainly driven by gravitational effects associated to cosmic structure formation, is very similar in the three sets of simulations (see Appendix \ref{app:physics}).

\item From the analysis of the shock cell distribution function around one of the most massive clusters in our sample, we find that AGN feedback tends to produce a small shortfall of shocked cells in the range $4\mincir\mathcal{M}\mincir60$, compared to the {\tt \nr} run.
This discrepancy is similarly present when comparing the {\tt \csf} and the {\tt \nr} simulation, suggesting its origin to be related to the impact of radiative processes in general (see Appendix \ref{app:physics}).

\end{itemize}

}
 
Although shocks can be observed at different wavelengths, such as X-rays, millimetric (via the SZ effect) and radio, their detection is challenging. However, the key role played by shocks in the thermalization and energetics of the LSS makes mandatory observing and understanding them in detail. In this regard,  it is crucial to perform detailed analyses of the formation and evolution of shocks in {\it realistic} and complex simulations like the ones analyzed in this project. These studies need however to be improved in terms of shock identification and estimation of the energy dissipation at shocks. Moreover, even when the analyzed cluster sample is perfectly suitable for the present study,  larger samples are needed in order to reach more robust statistical conclusions. The combination of results from simulations with the expected observations from the next generation of X-rays, SZ and radio instruments \citep[see, for instance,][]{Vazza_2019} will be essential to deepen our understanding of the main properties of the LSS of the Universe.

\section*{ACKNOWLEDGEMENTS}

We thank the anonymous referee for his/her constructive comments.
This work has been performed under the Project HPC-EUROPA3 (INFRAIA-2016-1-730897), with the support of the EC Research Innovation Action under the H2020 Programme; in particular, SP gratefully acknowledges the hospitality of the Department of Physics of the University of Trieste and the computer resources and technical support provided by CINECA.
SP and VQ acknowledge financial support from the {\it Spanish Mi\/nisterio de Ciencia, Innovaci\'on y Universidades} (MICINN, grant PID2019-107427GB-C33)  and the Ge\/neralitat Valenciana (grant PROMETEO/2019/071). SB and ER acknowledge financial support from the PRIN-MIUR 2015W7KAWC grant and the INFN InDARK grant.
VB acknowledges support by the DFG project nr. 415510302.
KD acknowledges support by the Deutsche Forschungsgemeinschaft (DFG, German Research Foundation) under Germany's Excellence
Strategy - EXC-2094 - 3907833 as well as the COMPLEX project from the European Research Council (ERC) under the European
Union's Horizon 2020 research and innovation program grant agreement ERC-2019-AdG 860744.

\section*{Data availability}
The data underlying this article will be shared on reasonable request to the corresponding author.

\bibliographystyle{mnbst}
\bibliography{Mach_paper_v3}


\appendix

\section{Dependence on grid resolution}
\label{app:resol}

\begin{figure}
\begin{center}
{\includegraphics[width=7.5cm]{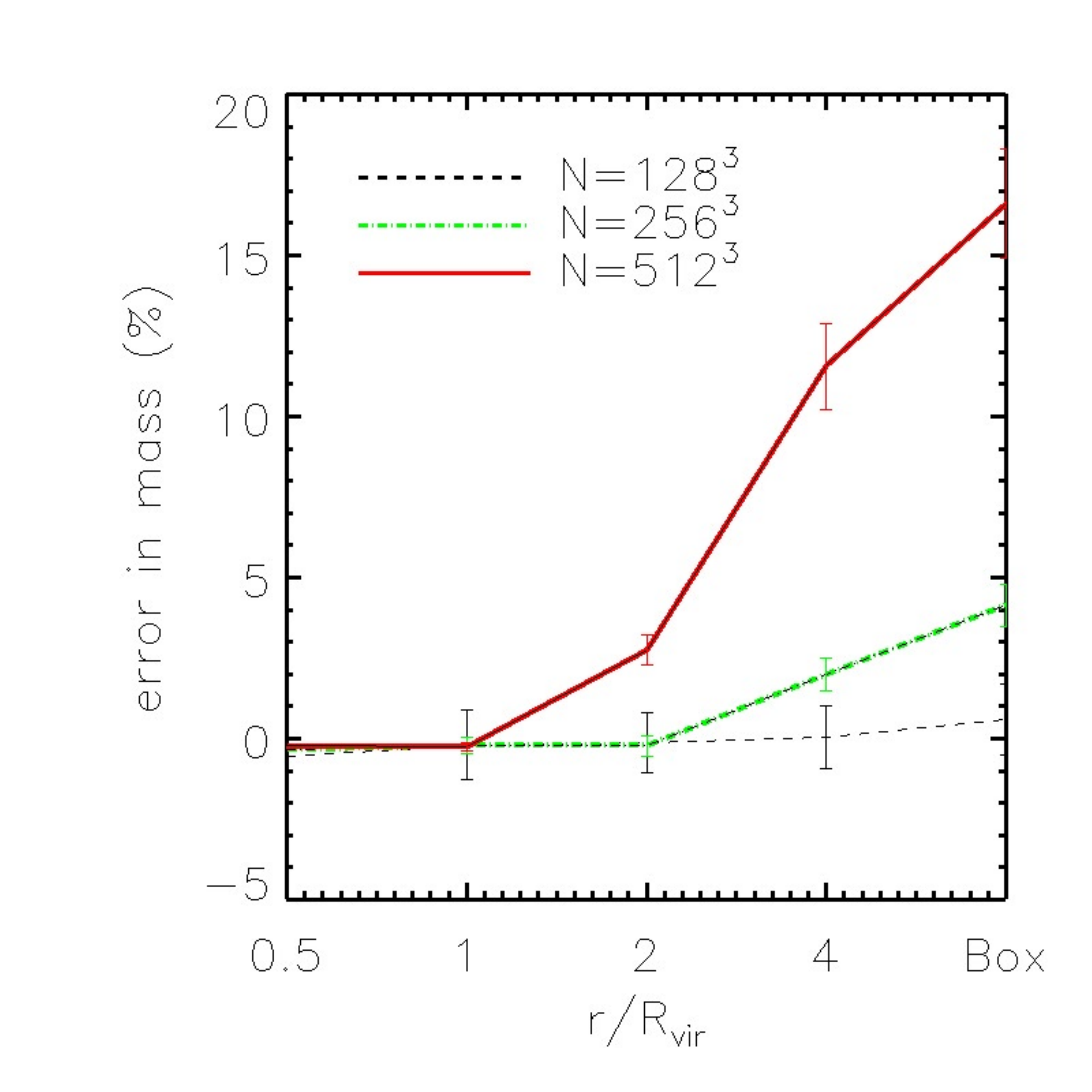}}
\caption{{Mean relative error in recovering the gas mass within different radial apertures from the cluster centre as obtained for the whole sample of 29 clusters in the {\tt \agn} simulations at $z=0$. The relative error in mass is computed as $(M_p-M_{grid})/M_{p}$, being $M_p$ and $M_{grid}$ the masses obtained from the distribution of particles and from the integral of the gas density over the grid. 
The considered radial apertures are spherical regions of radius equal to 0.5, 1, 2 and 4 times the virial radius of each halo. The last tick mark on the x-axis stands for the volume of the whole cubical box with side length equal to $8\times R_{vir}$. Results are shown for the resolutions corresponding to discretizing the considered volumes with  $128^3$, $256^3$ and $512^3$ cells (black, green and red lines, respectively). Error bars stand for $1-\sigma$ standard deviation around each mean distribution. }}
\label{fig:mass_error_resol}
\end{center}
\end{figure}

\begin{figure}
\begin{center}
{\includegraphics[width=7.5cm]{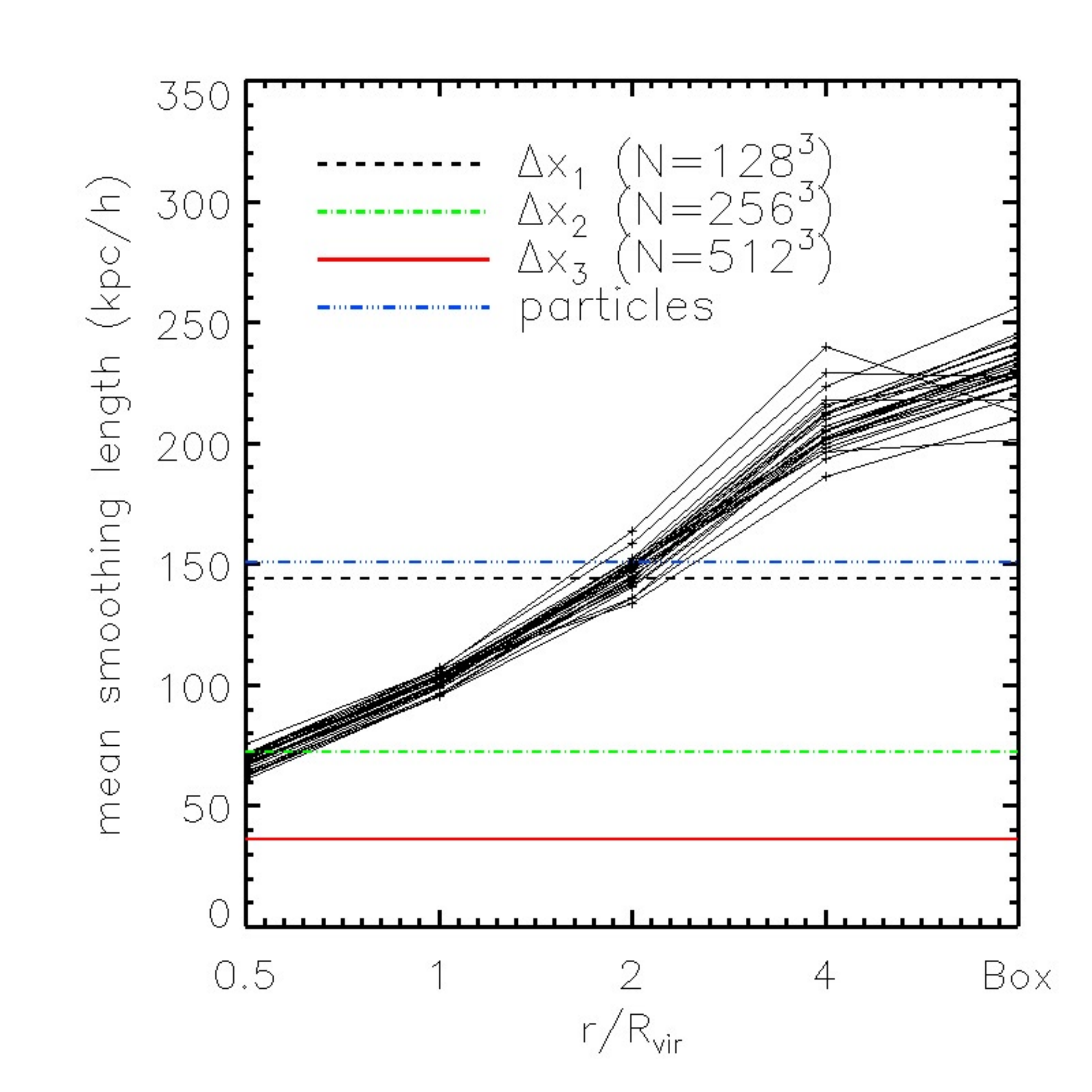}}
\caption{{Distribution of the mean smoothing length of all the gas particles within different radial apertures from the cluster centre for each of the 29 main halos in the {\tt \agn} simulations at $z=0$ (black continuous lines). The considered radial apertures are the same as those shown in Fig.~\ref{fig:mass_error_resol}.
Horizontal lines represent the corresponding mean grid resolutions when discretizing the considered volumes with  $128^3$, $256^3$ and $512^3$ cells (horizontal black, green and red lines, respectively). An additional horizontal blue line represents the mean smoothing length 
of all the gas particles within the whole simulation as obtained from the whole sample of 29 clusters.}}
\label{fig:hs_mean_resol}
\end{center}
\end{figure}

When we smooth a distribution of SPH particles onto a grid, ideally, we want to reach a grid resolution comparable with the resolution we have in the SPH simulation. However, this is not trivial, especially when we employ a regular grid \citep[see][for a recent analysis of different smoothing methods]{Rottgers_2018}.  
Therefore, once we have smoothed a distribution of particles onto a grid, we need to make sure that, for instance, the total mass inside the considered volume is  conserved. This means that we should get the same gas mass within a given region when comparing the summation over all SPH particles with the integral of the gas density over a grid. To address this issue, Fig.~\ref{fig:mass_error_resol}  shows the mean relative error in  recovering the gas mass within different radial apertures from the cluster centre as obtained for the whole sample of 29 clusters in the {\tt \agn} simulations at $z=0$. The considered apertures are spherical regions of radius equal to $0.5, 1, 2$ and 4 times the virial radius of each halo. 
The last tick mark on the x-axis, labelled as `Box', stands for the volume of the whole cubical box, that is, including as well the corners of the box and not just the sphere, with side length equal to $8\times R_{vir}$. Then, the error in mass is computed as the relative difference between the mass obtained from the direct summation over the sample of particles within a given region and from the integration of the gas density smoothed out onto a grid of a given resolution, that is, $(M_p-M_{grid})/M_{p}$, being $M_p$ and $M_{grid}$ the masses derived from the distribution of particles and from the grid. Results are shown for three different resolutions which correspond to discretize the considered volumes with $128^3$, $256^3$ and $512^3$ cells (black, green and red lines, respectively). The results shown in this figure  indicate that when we consider the lowest resolution, the one corresponding to $128^3$ cells, the error in mass is around $\sim$1 per cent for all  the considered radial apertures. Instead, when we increase the grid resolution the error increases significantly, especially for the resolution corresponding to $512^3$ cells and for radial apertures beyond $2\times R_{vir}$.

To understand better this trend, Fig.~\ref{fig:hs_mean_resol} shows the distribution of the mean smoothing length of all the gas particles within different radial apertures from the cluster centre for each of the 29 main halos in the {\tt \agn} simulations at $z=0$ (black continuous lines). As expected, the mean smoothing length of particles within clusters is relatively low within the virial radius ($\sim70\, h^{-1}$kpc) and increases in outer regions ($\sim220 \,h^{-1}$kpc). As in the previous figure, we have considered spherical regions of radius equal to $0.5, 1, 2$ and 4 times the virial radius of each halo. We have also considered the volume of the whole cubical box with side length equal to $8\times R_{vir}$. Horizontal lines represent the corresponding mean grid resolutions when discretizing the considered volumes with  $128^3$, $256^3$ and $512^3$ cells (horizontal black, green and red lines, respectively). An additional horizontal blue line represents the mean smoothing length of all gas particles as obtained from the whole sample of 29 clusters.
According to these results, the resolution corresponding to a grid with $128^3$ cells (horizontal black line) is comparable to the mean smoothing length of all particles ($\sim 150\, h^{-1}$kpc; horizontal blue line).  Although this is a simple approximation, in view of the results shown in Figs.~\ref{fig:mass_error_resol} and \ref{fig:hs_mean_resol}, we decided to adopt as our reference resolution the one corresponding to discretize the distribution of particles onto a regular grid with $128^3$ cubical cells.

Early studies with grid-based cosmological codes \citep[e.g.][]{Kang_2007} have shown that for a fixed grid resolution it is difficult to capture smaller-scale shock features, meaning that shocks in inner cluster regions may not be fully identified. In our study we have checked that, as the spatial grid resolution is artificially increased, a larger amount of weaker and thinner shock surfaces can be identified. We should, however, take this result with caution since, {as shown for the mass conservation in Fig.~\ref{fig:mass_error_resol},} the artificial improvement of the grid resolution can generate unrealistic and spurious identification of shocks. 
Therefore, trying to resolve shocks in cluster outskirts with a grid resolution higher than the actual SPH resolution may introduce significant spurious effects. Indeed, the proper way to study the effect of spatial resolution on our results would be to re-simulate the sample of 29 regions, with the same initial conditions, but with different mass resolutions. 
Although this option is beyond the purpose of this paper, it could be addressed in future works.
Therefore, taking into account the results discussed in this section, we have chosen the resolution corresponding to $N=128^3$ cells as our reference one throughout the paper. 

\section{Connection between shocks and galaxy clusters}
\label{app:clusters_and_shocks}

\begin{table*}
  \begin{tabular}{lcccccccc}
  \hline
  cluster & $\mvir$ & $\rvir$ & $\rfive$ & core & dynamical &  $\Delta x_1$ & $\Delta x_2$ & $\Delta x_3$ \\
   &  [$10^{14}\msunh$] & [$\mpch$] & [$\mpch$] & state & state &  [$\kpch$] & [$\kpch$] & [$\kpch$] \\
  \hline
  D2  & 5.26  & 1.70 & 0.74 & CC  &  Intermediate & 105.3 & 52.9 & 26.5  \\
  D3  & 6.53  & 1.83 & 0.84 & NCC &  Intermediate & 113.3 & 56.9 & 28.5 \\
  D4  &  5.21  & 1.69 & 0.73 &  NCC     & Disturbed & 105.0 & 52.7 & 26.4  \\
  D6  & 15.41 & 2.43 & 1.08 & NCC & Intermediate & 150.8 & 75.7 & 37.9  \\
  D10 & 15.46 & 2.43 & 1.07 & CC  &  Disturbed & 151.1 & 75.9 & 38.0  \\
  D21 &  14.26 & 2.37   & 1.17   & CC      & Regular & 146.9 & 73.8 & 37.0  \\
\hline
\end{tabular}
\caption{Main properties of the six selected clusters at $z=0$ in the {\tt \agn} run. Columns from
  left to right give, respectively, the cluster id, the viral mass
  ($\mvir$), the virial ($\rvir$) and $\rfive$ radii, the CC/NCC classification based on their core properties, the regular/disturbed/intermediate classification based on their global dynamical state, and the spatial resolution of the computational boxes built around them when we consider a box of side length equal to $8\times R_{vir}$ discretized with a number of cells equal to $N_1=128^3$, $N_2=256^3$, $N_3=512^3$.}
\label{tab:clusters}
\end{table*}

In order to analyze in more detail the connection between the distribution of shock waves and  the dynamical or cool-coreness state of the clusters, in this Section we will focus our analysis on a smaller selection of 6 systems, whose main properties are summarized in Table \ref{tab:clusters}. Although this reduced sample is not statistically representative, it has been chosen in order to have systems with different masses, core and dynamical classifications.

\begin{figure*}
\begin{center}
{\includegraphics[width=15.5cm]{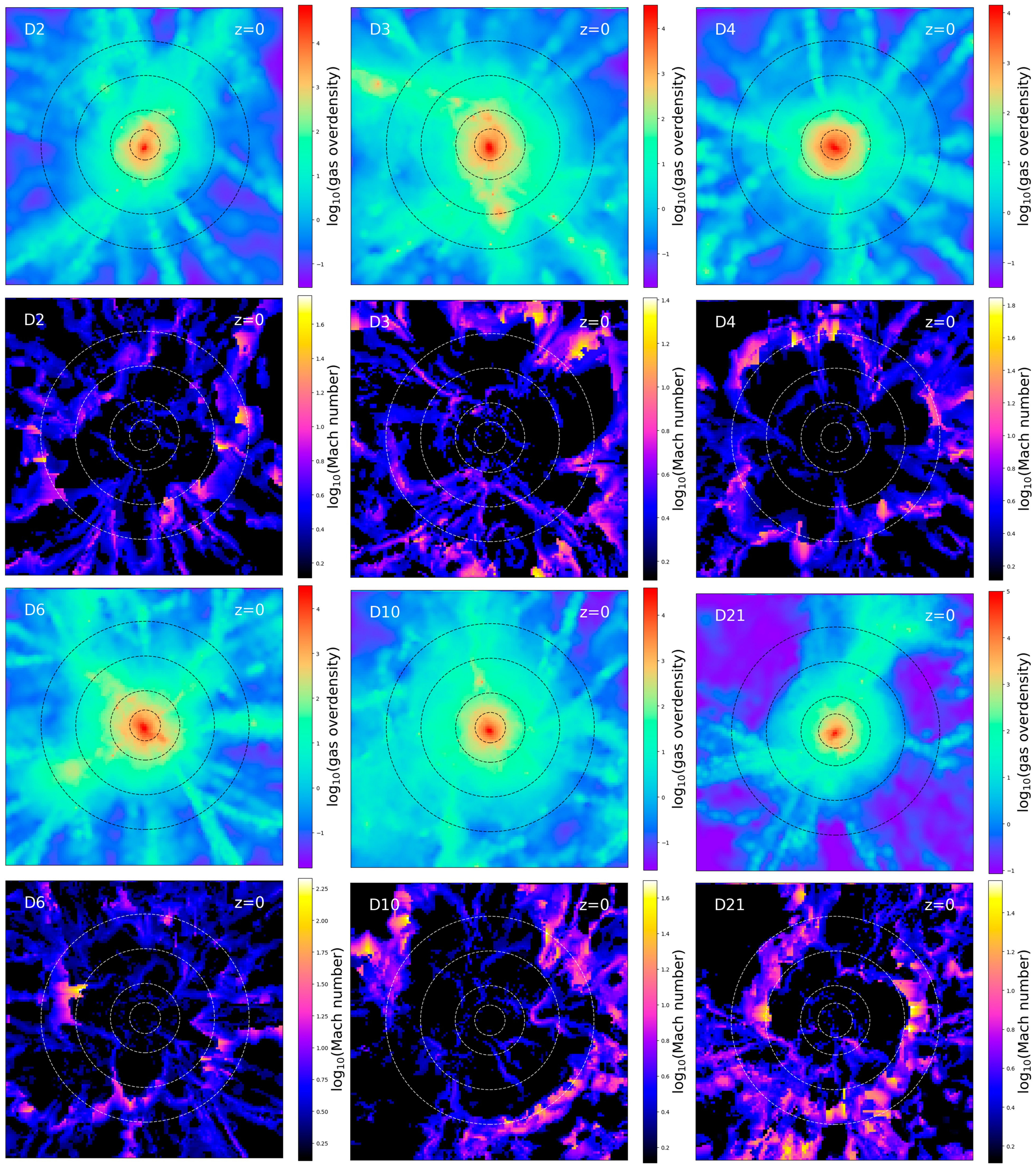}}
\caption{2-D projections of the gas overdensity and the Mach number distribution at $z=0$ around each of the 6 clusters  whose properties are given in Table \ref{tab:clusters}. Each map, projected along the z-axis, has a side length of $8\times R_{vir}$ and a projection depth of 7 cells centred on the cluster position. Circles on the maps represent, respectively, $R_{500}$, $R_{vir}$, $2\times R_{vir}$ and $3\times R_{vir}$  of the central object.}
\label{fig:only_dens}
\end{center}
\end{figure*}

\begin{figure*}
\begin{center}
{\includegraphics[width=7.5cm]{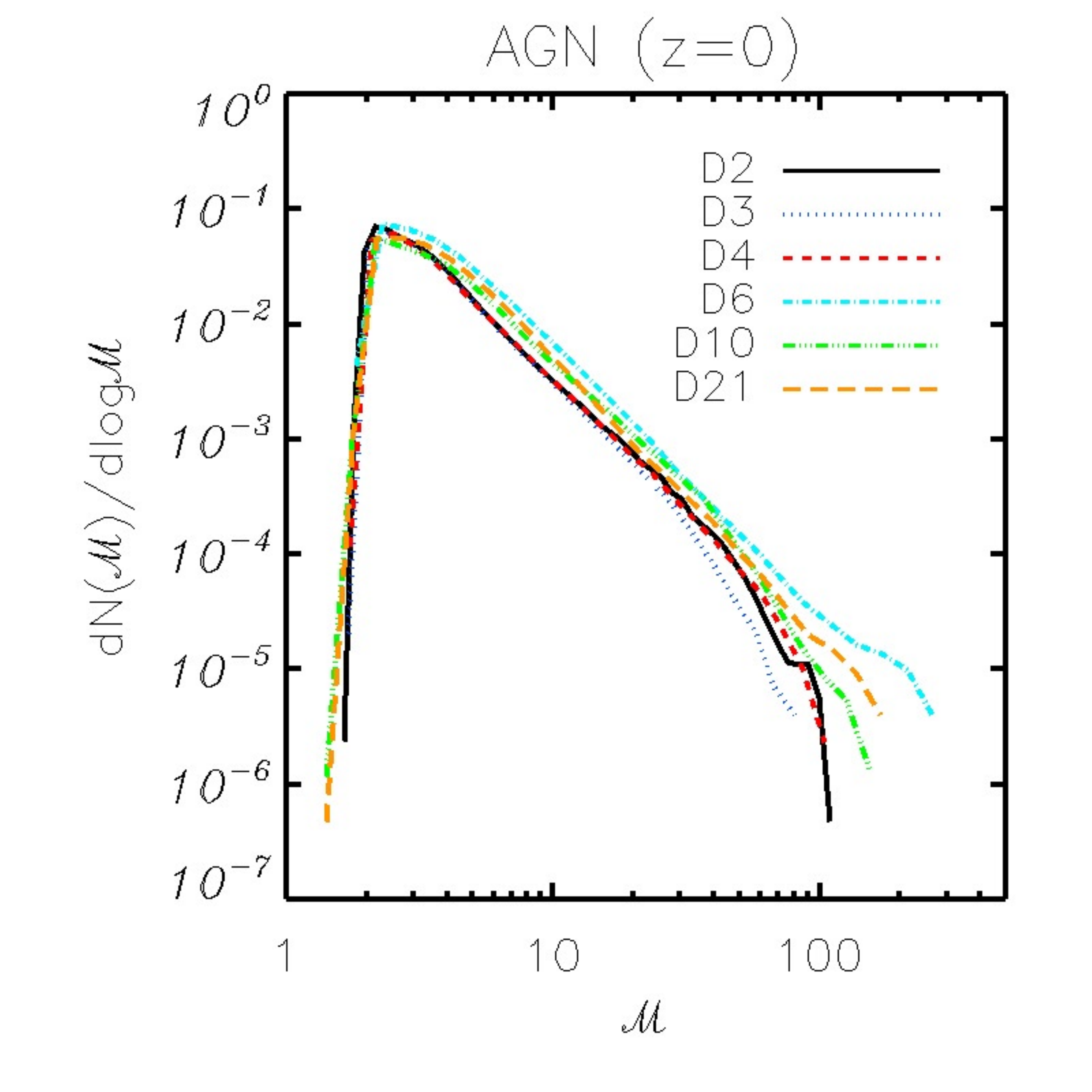}
\includegraphics[width=7.5cm]{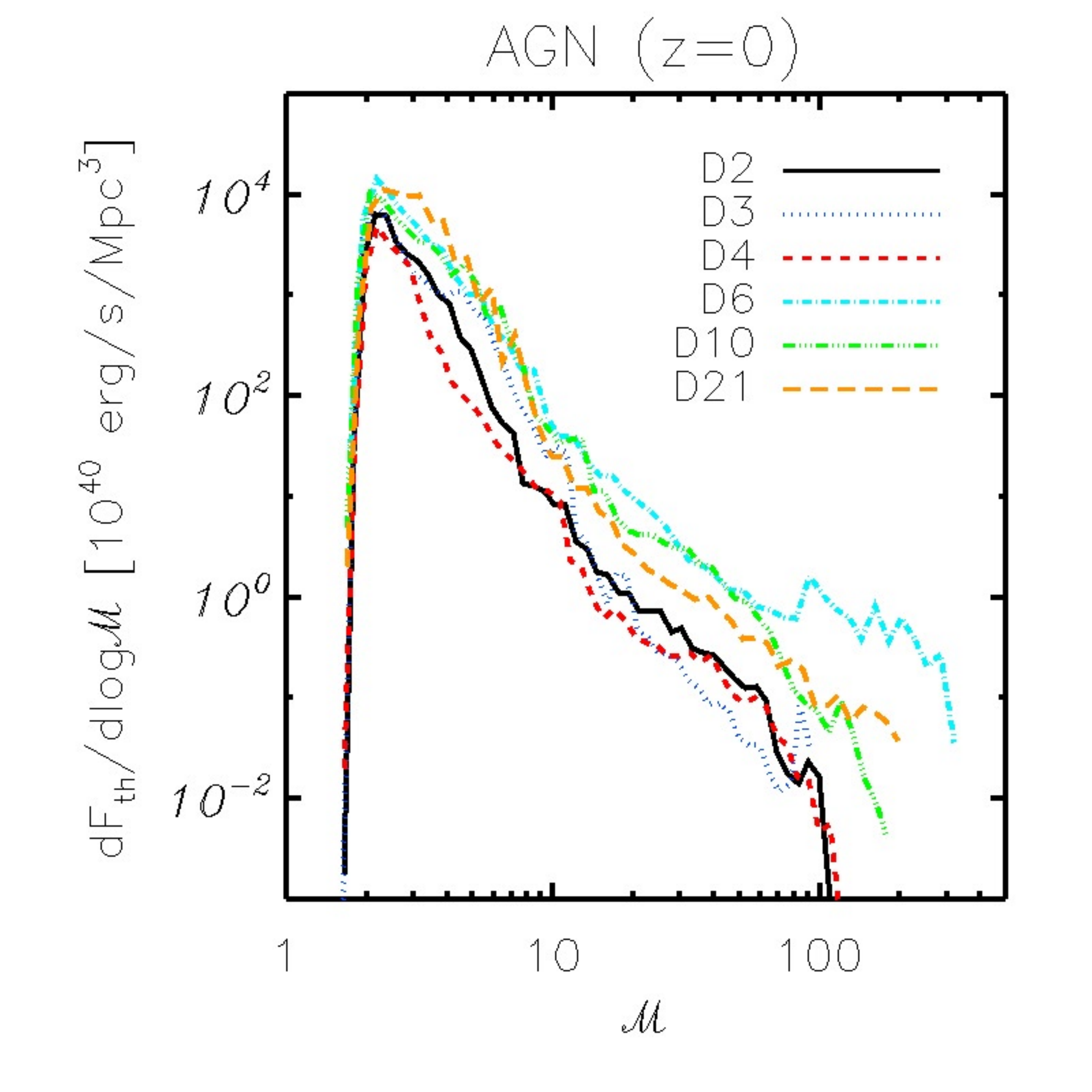}}
\caption{{\it Left panel:}  Shock cell distribution function for the reduced sample of clusters in the  {\tt \agn} simulation at $z=0$. {\it Right panel:} Differential distribution of the thermal energy through shocks (divided by the volume of each region) for the reduced sample of clusters in the  {\tt \agn} simulation at $z=0$.}
\label{fig:distri_selection}
\end{center}
\end{figure*}

\begin{figure*}
\begin{center}
{\includegraphics[width=15cm]{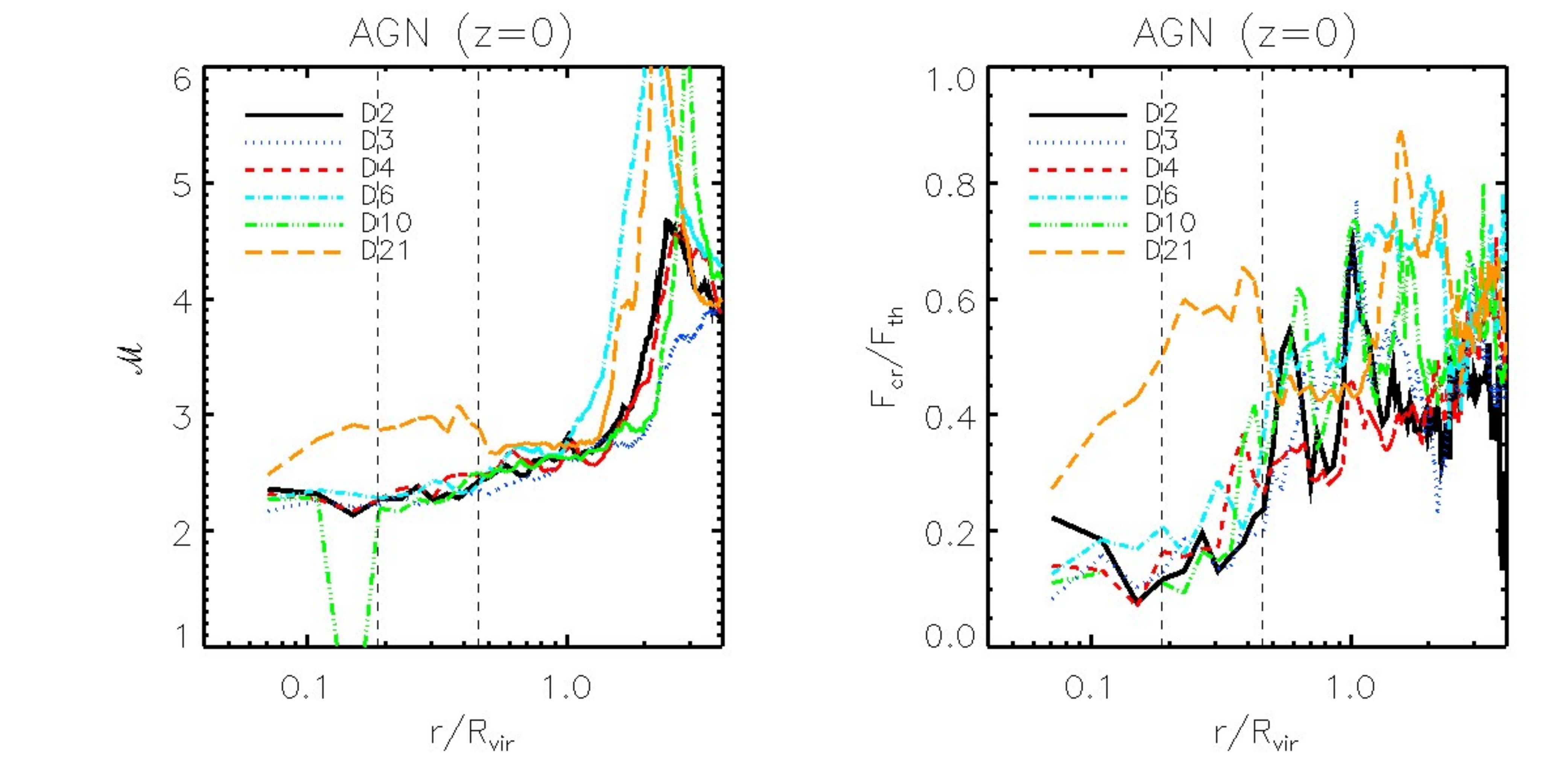}}
\caption{{\it Left panel:} Volume-averaged radial profiles  of the mean Mach number for the six main central clusters of regions D2, D3, D4, D6, D10 and D21 out to $4\times R_{vir}$. 
{\it Right panel:} Mean radial profiles as in the left panel but for the ratio of the CR energy flux over the thermal energy flux out to $4\times R_{vir}$ of each cluster. Vertical lines mark the position of the mean value of radii $R_{2500}$ and $R_{500}$.}
\label{fig:radial_prof}
\end{center}
\end{figure*}

\begin{figure}
\begin{center}
{\includegraphics[width=7.5cm]{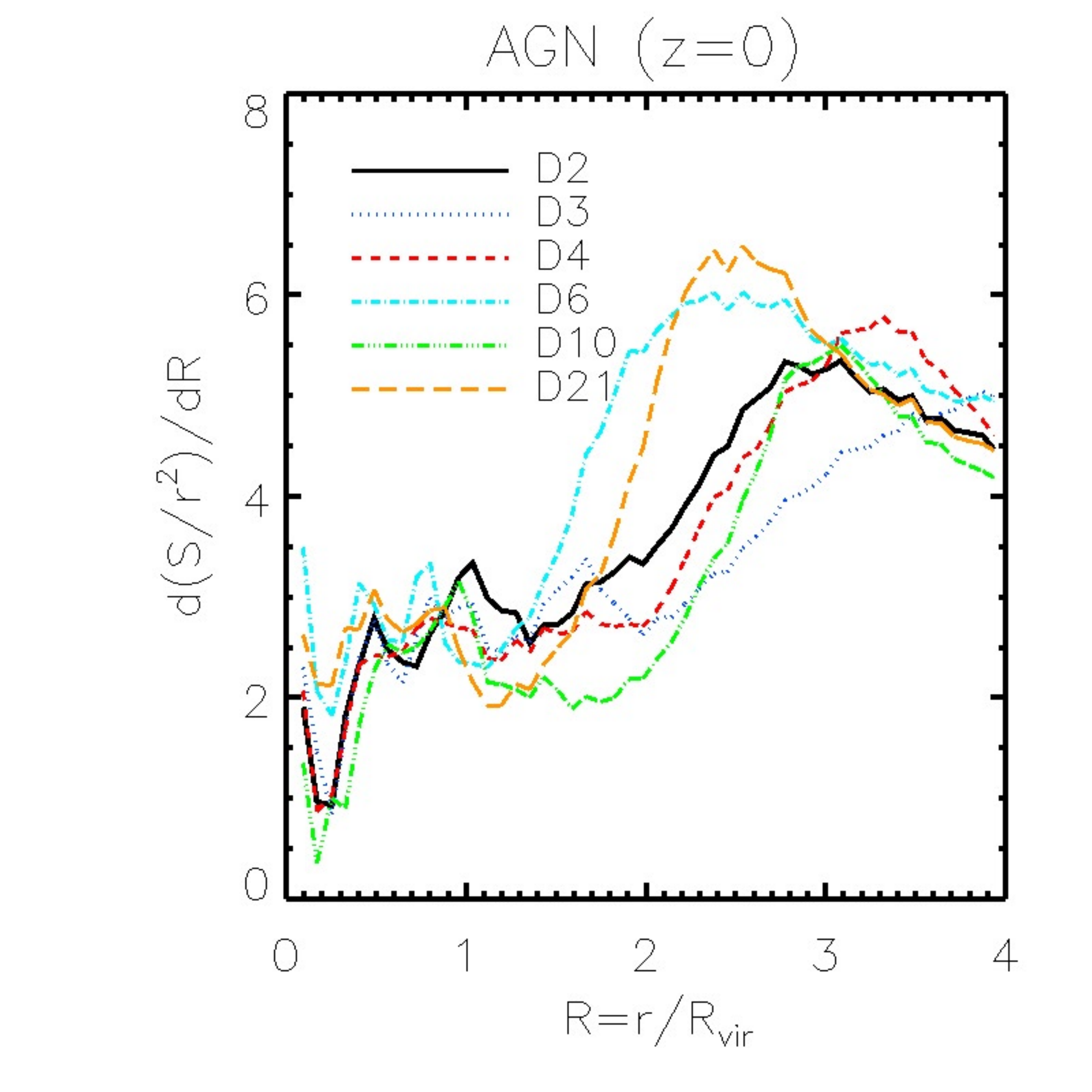}}
\caption{Radial shock surface distribution out to $4\times R_{vir}$ for the reduced sample of clusters in the {\tt \agn} run at $z=0$.}
\label{fig:radial_prof_ssd}
\end{center}
\end{figure}

\begin{figure*}
{\includegraphics[width=6.2cm]{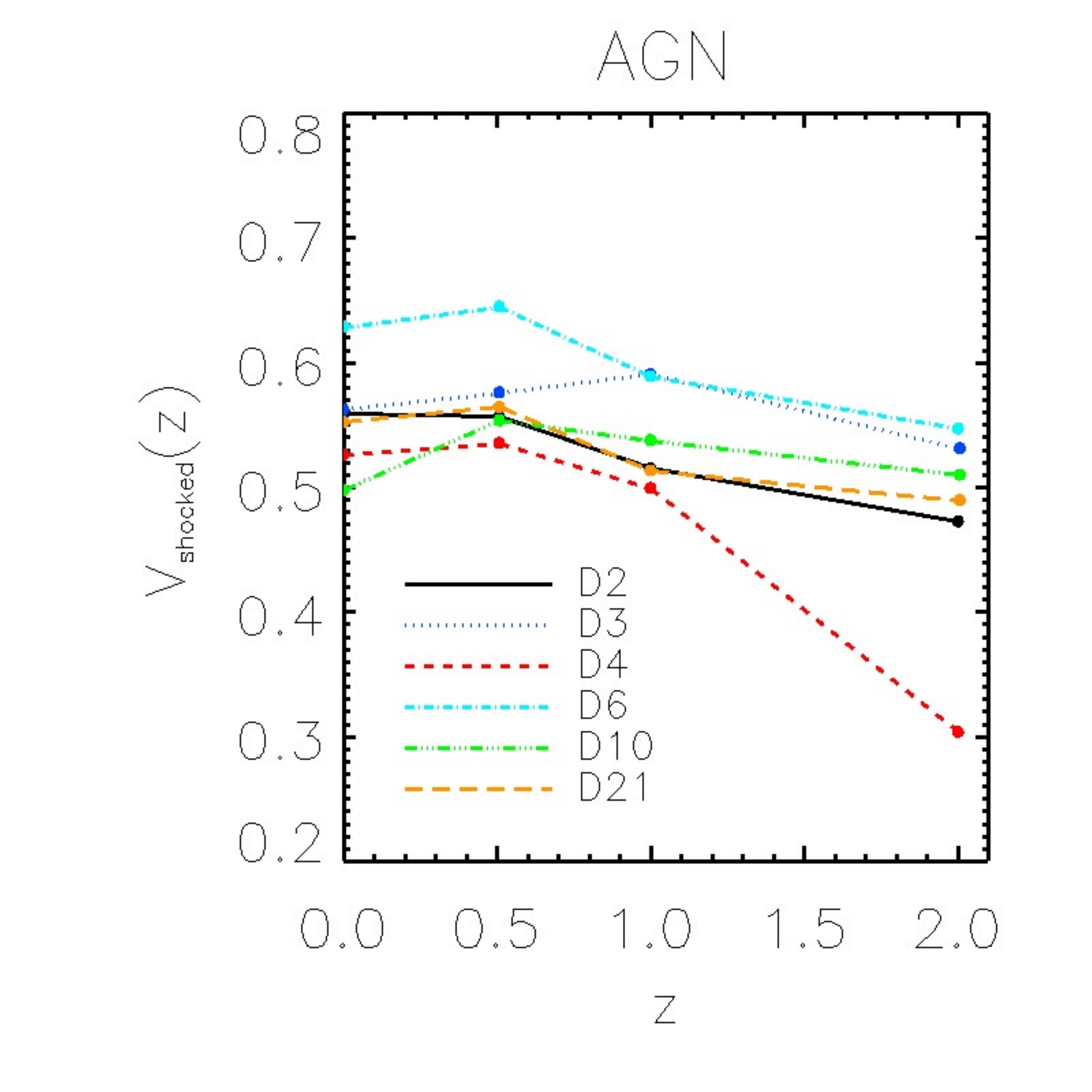}
\hspace{-0.7cm}
\includegraphics[width=6.2cm]{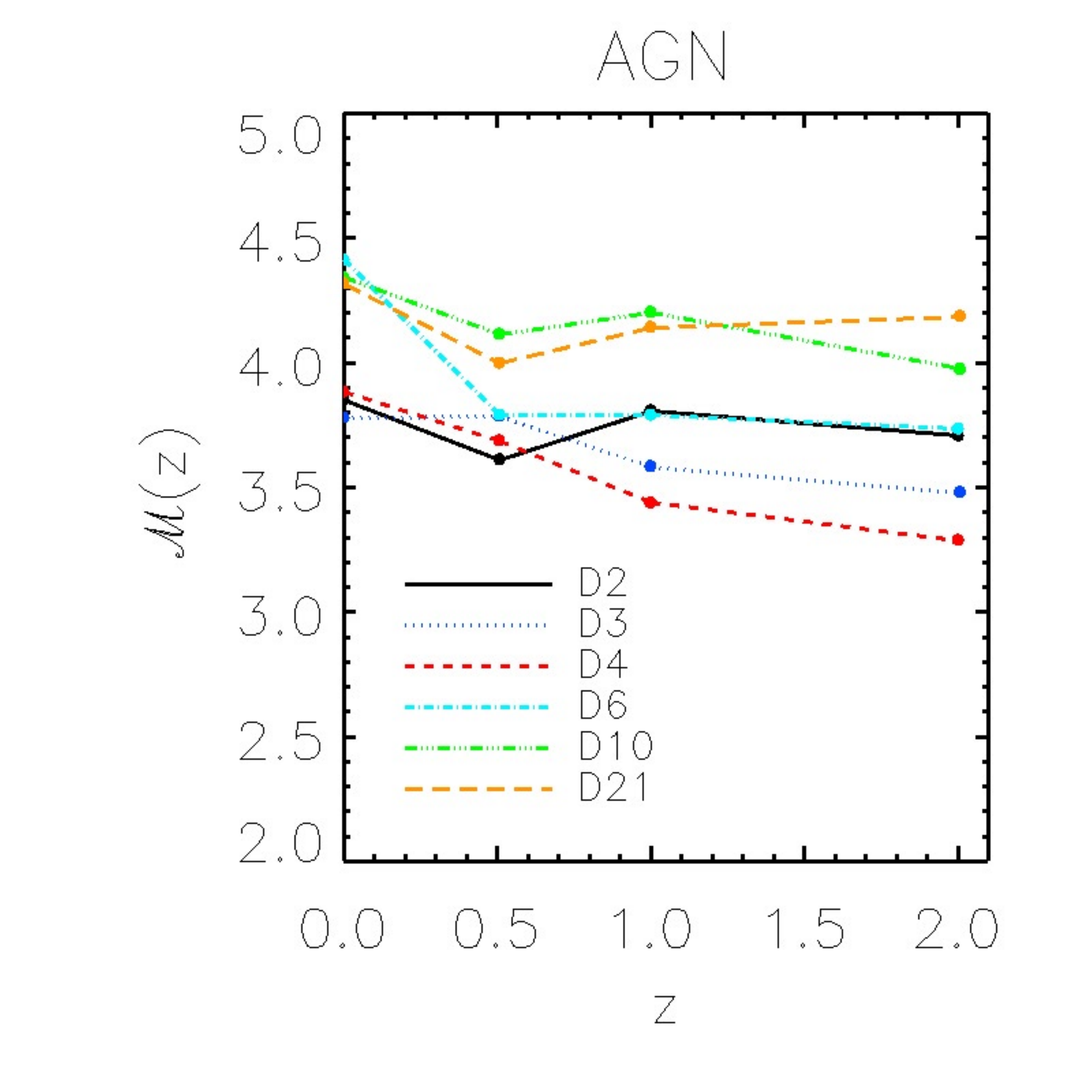}
\hspace{-0.7cm}
\includegraphics[width=6.2cm]{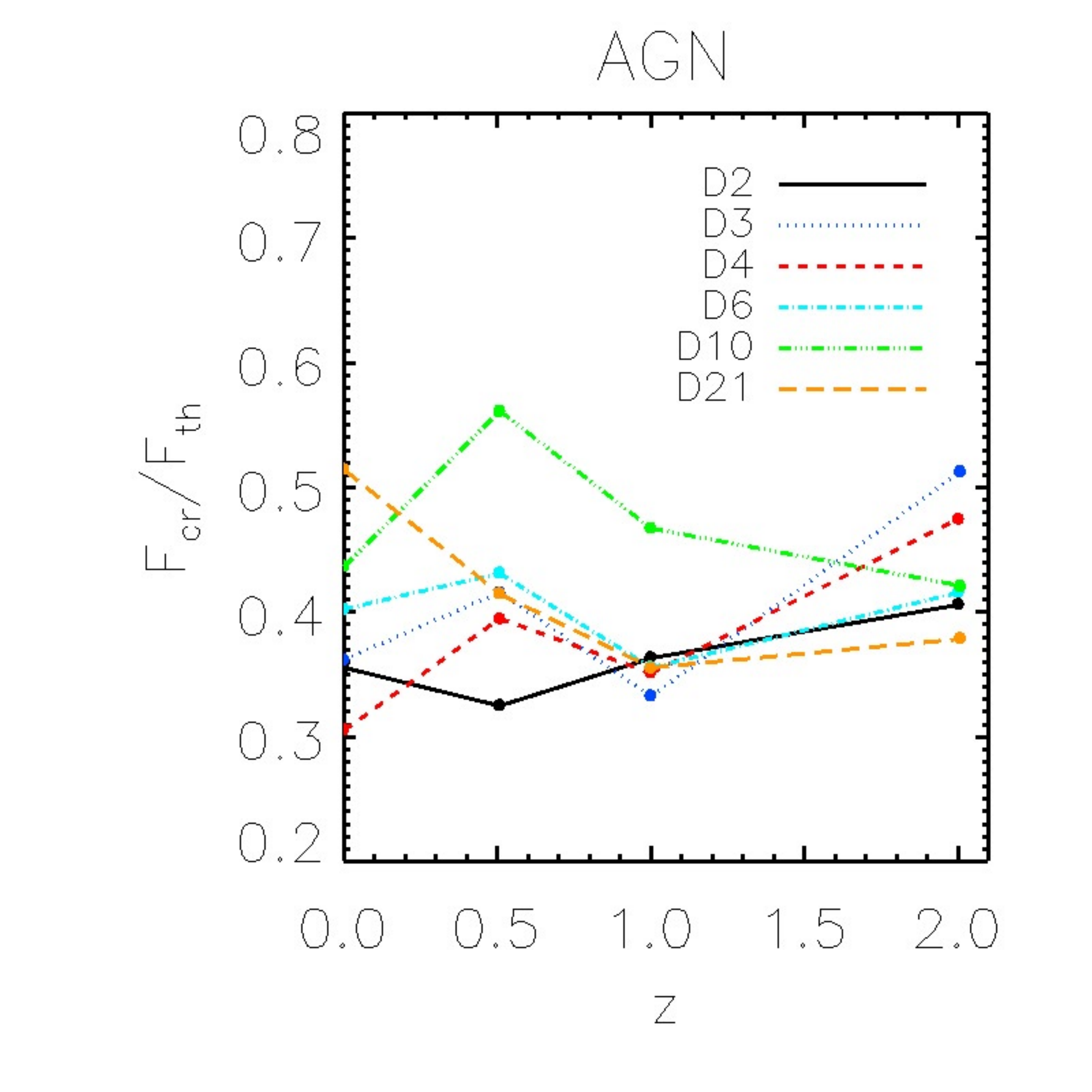}}
\caption{Redshift evolution of different global shock related quantities within the computational domains associated to the six selected regions D2, D3, D4, D6, D10 and D21. From left to right: (i) redshift evolution of the fraction of the computational volume hosting shocked cells; (ii) redshift evolution of the mean Mach number of all the shocked cells; and (iii) redshift evolution of the ratio of CR energy flux over the thermal energy flux through shocked cells in each computational volume. }
\label{fig:z_evol}
\end{figure*}

Figure \ref{fig:only_dens} shows projections along the z-axis of the gas overdensity and the Mach number distribution around each of the 6 selected cluster regions  at $z=0$ (including region D21, whose redshift evolution has been shown in Fig.~\ref{fig:z_evol_maps}). For the central cluster of each region, radii $R_{500}$, $R_{vir}$, $2\times R_{vir}$ and $3\times R_{vir}$ are also shown. 
According to the density maps, these projections clearly show a wide variety of large-scale structures around the different haloes. Indeed, outside the virial radius, a complex and irregular distribution of filaments and void regions dominate the simulated volume. As expected, within the clusters' virial region the density distribution is dominated by the very central area within $R_{500}$, although smaller substructures are also visible in outer regions, in-between  $R_{500}$ and $R_{vir}$.

Interestingly, if we look instead at the shock Mach number maps shown in Fig.~\ref{fig:only_dens}, most of the regions show a high-Mach number shock surface (located at several virial radii from the cluster centre) wrapping the large scale density distribution {and separating the external unshocked gas from the cluster outskirts}. {Although the shape of these external accretion shocks is assumed to be  quasi-spherical, as in region $D21$, there are some other regions, like $D3$, $D4$ or $D6$, which show a more irregular ``flower-like'' distribution. Some studies have shown that this particular pattern is quite common and results from the intersection between merger and accretion shocks developed during clusters' evolution \citep[e.g.][]{Zhang_2020}.} Instead, within the virial radius clusters show different and more irregular shock patterns. In some regions, like $D2$ or $D10$, some bow shocks with low Mach number are detectable in the area within $R_{500}$ and  $R_{vir}$.

Similarly to Fig.~\ref{fig:distri_all_halos}, we show in Fig.~\ref{fig:distri_selection} the shock cell distribution function (left panel) and the differential distribution of the thermal energy flux through shocks (right panel) for the reduced sample of clusters in the  {\tt \agn} simulation at $z=0$. These distributions have been computed for a cubical region of side length equal to $8\times R_{vir}$ around each cluster. {Obviously,} for all clusters both distributions show a broad agreement with the mean trend shown in Section \ref{sec:shock_distribution} for the complete sample of systems. In general, we do not distinguish any particular trend depending on the core or global dynamical state of the different clusters. However, for a fixed Mach number, the most massive clusters (i.e. D6, D10 and D21) tend to host a larger percentage of shocked cells and seem to be more efficient in thermalizing the IGM, in agreement with the results shown in Fig.~\ref{fig:corr_all_halos}.  
In addition, even though regions D6, D10 and D21 are quite similar in terms of the shock cell distribution function through most of the Mach number range, shocks within region D21 (connected to the only dynamically regular system in our sample) seem to process a lower amount of energy than shocks within regions D6 and D10, especially for $10\mincir \mathcal{M} \mincir 80$. A similar trend was also obtained by \citet{Vazza_2010}, who found mild differences, especially  for shocks with $\mathcal{M}>10$, between their merging and relaxed cluster samples, with the relaxed systems processing $\sim10$ times less thermal energy than the merging ones.  

It is also important to stress that the distributions shown in Figs.~\ref{fig:distri_all_halos} and \ref{fig:distri_selection} are derived from the 3D distribution of shocks around each cluster, while the maps are projections of a slice across the cluster centre. In this way, most of the shocks that boost the SDF at high Mach number for region D6 are not visible from the map.

The left panel of Fig.~\ref{fig:radial_prof} reports the mean Mach number radial profiles, out to $4\times R_{vir}$, for the reduced sample of clusters at $z=0$. Despite the different dynamical classification of the systems, they show very similar radial Mach number distributions, especially within $R_{500}$. Indeed,  out to the viral radius, all clusters show mean Mach number values within $2\mincir \mathcal{M} \mincir 3$. Above $R_{vir}$ the mean strength of shocks smoothly increases, reaching values as high as $\mathcal{M}\mincir 4$ at $2\times R_{vir}$. 
As discussed in Section \ref{sec:shock_distribution}, the shape of these profiles is in line with the results presented in previous analyses  \citep[e.g.][]{Vazza_2009b,Vazza_2010,Vazza_2011b}. In our case, we do not find either a clear trend depending on the dynamical state of the clusters nor on their cool-coreness. In this regard, it is interesting to remind that the classification in relaxed/disturbed systems was done within $R_{200}$ and, therefore, it does not mean that a relaxed cluster must have a smooth gas density distribution up to $\sim2-3\, R_{vir}$.
{However, we still note that at $\sim2-3\, R_{vir}$ the shock cell distributions within regions D6, D10 and D21 reach higher Mach numbers than shocks within the rest of regions.}

In order to explore the global efficiency of these systems in accelerating CR at shocks, the right panel of Fig.~\ref{fig:radial_prof} shows the mean radial profiles of the ratio between the CR energy flux and the thermal energy flux, $F_{CR}/F_{th}$, at $z=0$. On average, regardless of the dynamical state of the different clusters, we do not find relevant differences between them throughout the radial range. However, {region D21} shows  larger values than expected, especially below $R_{500}$.  These peaks in the $F_{CR}/F_{th}$ radial distribution must be connected with the slightly larger radial Mach number profiles shown by {this system} in the central regions (see left panel). 
Leaving aside these deviations, the mean radial distribution of the ratio $F_{CR}/F_{th}$ tends to increase with the distance from the cluster centre, providing {stringent} limits to the energy dissipation ratio. 
Although it is not possible to perform a straightforward comparison among different simulations, the obtained trend and  average value of the  $F_{CR}/F_{th}$ ratio is in line with the estimates shown in previous works \citep[see, e.g.,][where some dependences of the $F_{CR}/F_{th}$ ratio on the energy dissipation model employed are also shown]{Vazza_2009b}.

As it has been shown in Fig.~\ref{fig:radial_prof_all}, Fig.~\ref{fig:radial_prof_ssd} shows now the radial shock surface distribution, out to $4\times R_{vir}$, for the reduced sample of clusters in the {\tt \agn} run at $z=0$. Since this distribution gives an idea of the location of the main shock features within and around the cluster virial region, it can be correlated with the maps of the shock Mach number distribution shown in Fig.~\ref{fig:only_dens}. For each halo the bumps characterising the distribution within $R_{vir}$ highlight the presence of internal shock phenomena. Outside the virial radius, most clusters show a dominant peak in the radial shock surface distribution, which approximately indicates the position of the main external accretion shock. The distributions, however, show quite irregular shapes that depend on the particular clusters' environments and evolutionary histories. Although usually it is not possible to clearly distinguish the radius for the accretion shock (see, for instance, region D3), on average, we find the external accretion shock at a mean distance of $\sim2.5\times R_{vir}$ from the cluster centre, a value a bit larger than the results presented in previous studies \citep[e.g.][]{Schaal_2015, Schaal_2016,Nelson_2016}. As discussed in Section \ref{sec:shock_distribution}, the height of the external peaks in the radial shock surface distribution shows some dependence on cluster mass. In this case, leaving aside some unavoidable deviations resulting from the particular clusters' evolution, the height of the distributions is clearly larger in two of the most massive systems in our sample, i.e., $D6$ and $D21$. As for region $D10$, although the peak of the distribution is similar to that of the smallest systems, it starts from below and, therefore, its absolute increase is also remarkably higher than for regions $D2$, $D3$ or $D4$. This result is in line with the fact that, in general, the extension of accretion shocks is larger around the most massive clusters \citep[e.g.][]{Miniati_2000}.

We now investigate the temporal evolution of different global shock-related quantities for the reduced sample of clusters. Figure \ref{fig:z_evol}  shows the redshift evolution of the fraction of the computational volume hosting shocked cells (left panel), the  mean Mach number of all the cells hosting shocks (middle panel), and the ratio of CR energy flux over the thermal energy flux through shocked cells in each computational volume (right panel). All these quantities have been computed within a box of side length equal to $8\times R_{vir}$ of the main cluster of each selected region. From the analysis of this figure we obtain several interesting conclusions. 
The fraction of cells hosting shocks in the different simulated volumes slightly increases from {an average value of $\sim46$ per cent} at $z=2$ up to a value of $55-60$ per cent at $z=0$. The mean values obtained at $z=0$ are larger than those reported in previous studies \citep[$\sim20$\%, e.g.][]{Vazza_2009b, Planelles2013}. However, we should keep in mind that, in comparison with a cosmological volume, we are analysing a smaller and, therefore, a denser region around large clusters, which contributes to augment the amount of shocked cells. The mild increasing trend with decreasing redshift is in line with the results shown in Figs.~\ref{fig:z_evol_maps} and \ref{fig:distri_all_halos} for the complete sample of clusters as well as  with the results shown in other analyses \citep[e.g.][]{Vazza_2009b}. 
As for the z-evolution of the mean Mach number, the mean strength of shocks shows values within {$3\mincir\mathcal{M}\mincir 4$} for all clusters. In most of the regions the average Mach number tends to be slightly larger at $z=0$ than at $z=2$. In this regard, region $D21$, whose main central cluster was classified as a dynamically relaxed system, shows the lowest global variation between the high-z and the final Mach number values.       
According to the z-evolution of the $F_{CR}/F_{th}$ ratio, leaving aside some particular deviations, the mean amount of energy allocated for CR acceleration seems to be larger at higher redshift and to slightly decrease with time. In all regions, on average, the ratio of the CR over the thermal energy flux is {around} $\sim 0.4$. At the present epoch we find a value of $F_{CR}/F_{th}$ within {$\sim 0.3-0.5$}, in line with the values obtained in previous analyses \citep[e.g. within $\sim0.1-0.35$ in][depending on the adopted model of energy dissipation and on the considered environment]{Vazza_2009b}. 
Although the  approximation employed to estimate  $F_{CR}/F_{th}$ in post-processing is reasonable, we should keep in mind that this is a simplified treatment of the energy dissipation at shocks. Indeed, the fact that  $F_{CR}/F_{th}$ shows larger values at early cosmic epochs claims for the need to properly, and self-consistently, solve the dynamics of the CR population and to account for its effects on the thermal distribution of the IGM \citep[e.g.][]{Pfrommer_2006,Pfrommer_2007,Pfrommer_2008,Vazza_2016b}. This analysis is however beyond the scope of the present work.

\section{Dependence on baryonic physics}
\label{app:physics}

\begin{figure*}
{\includegraphics[width=15cm]{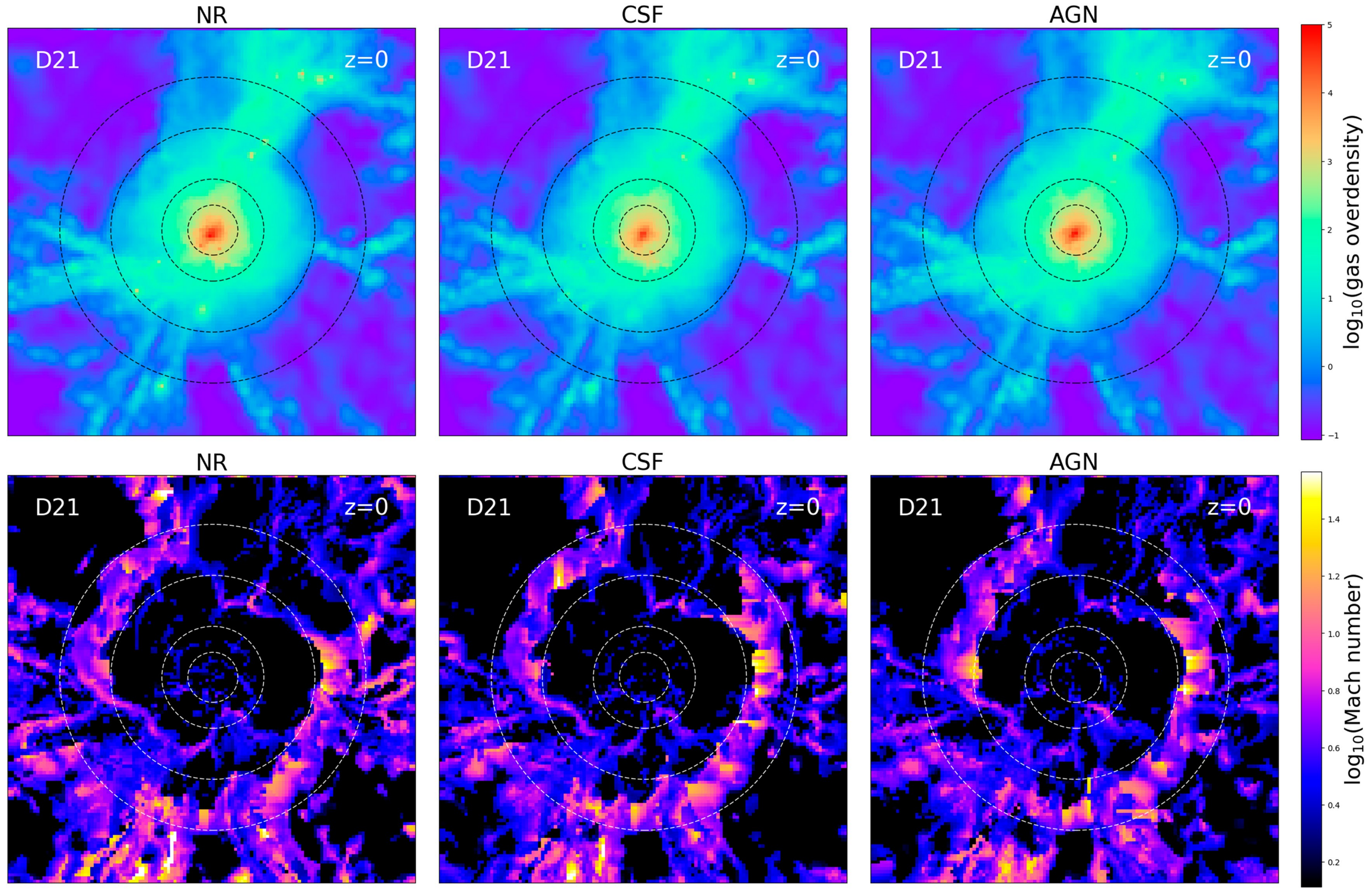}}
\caption{2-D zoom projections of the gas overdensity (upper panels) and the Mach number distribution (lower panels) around cluster D21 at $z=0$. Panels from left to right show the maps obtained for the {\tt \nr}, {\tt \csf} and {\tt \agn} simulations, respectively. Each map, projected along the z-axis, represents the whole region of side length equal to $8\times R_{vir}$. Circles on the maps represent $R_{500}$, $R_{vir}$, $2\times R_{vir}$ and $3\times R_{vir}$ of the central system.}
\label{fig:d21_physics_big_region}
\end{figure*}

\begin{figure*}
{\includegraphics[width=15cm]{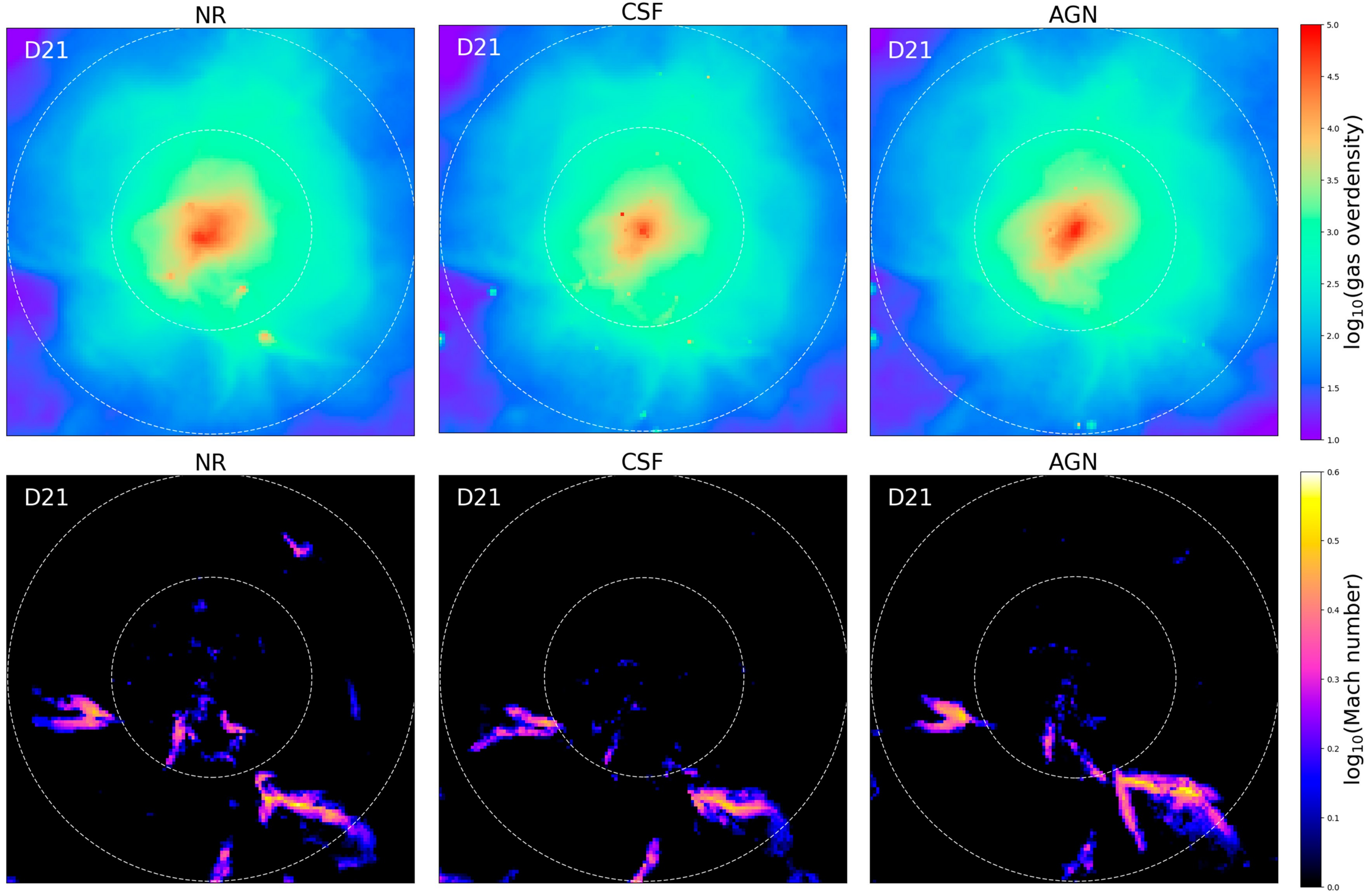}}
\caption{2-D zoom projections of the gas overdensity (upper panels) and the Mach number distribution (lower panels) around cluster D21 at $z=0$. Panels from left to right show the maps obtained for the {\tt \nr}, {\tt \csf} and {\tt \agn} simulations, respectively. Each map, projected along the z-axis, represents a zoom of  the cluster virial region. Circles on the maps represent $R_{500}$ and $R_{vir}$ of the central system. {These maps are obtained for a grid resolution corresponding to $512^3$ cells.}}
\label{fig:d21_physics}
\end{figure*}

\begin{figure*}
{\includegraphics[width=7.5cm]{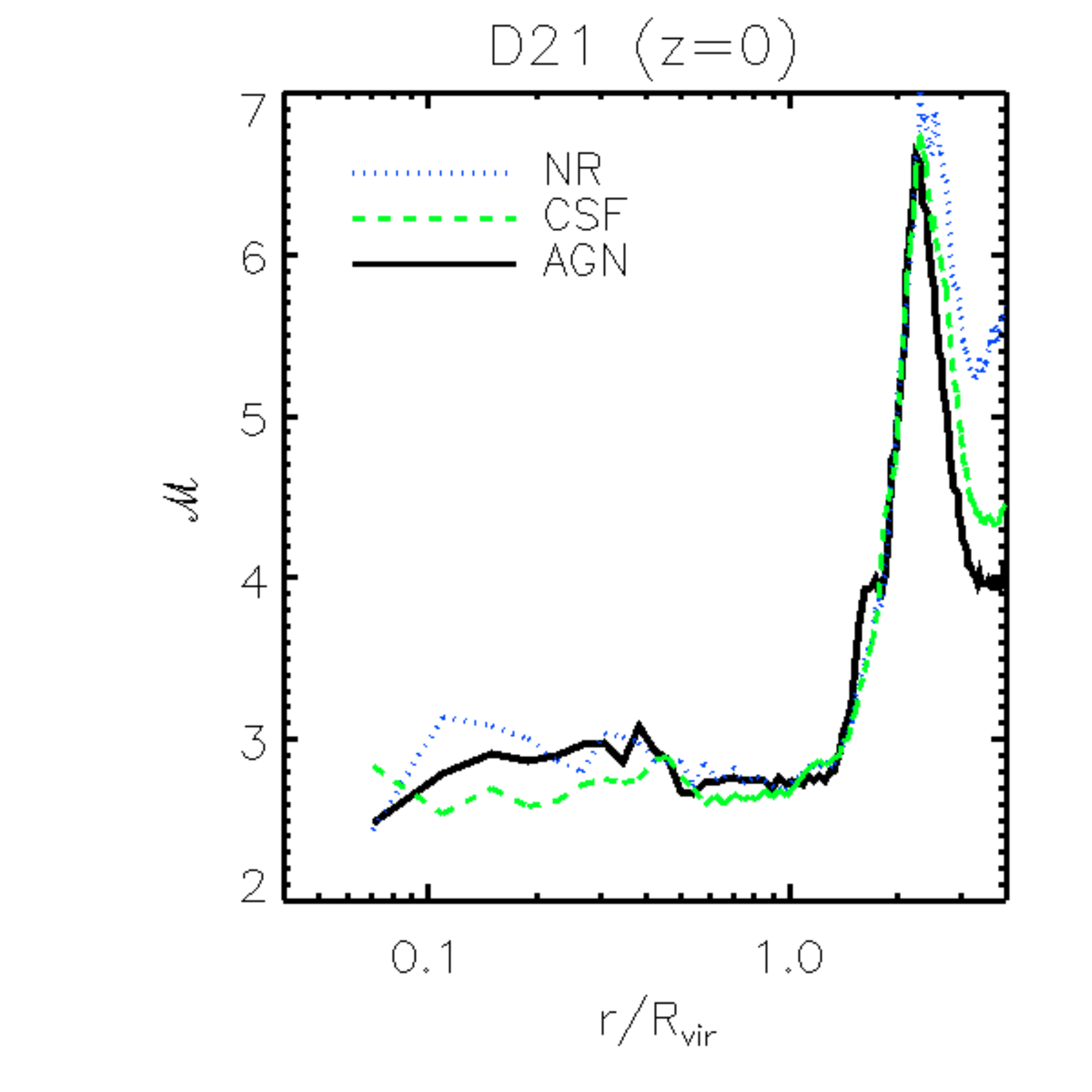}
\hspace{0.5cm}
\includegraphics[width=7.5cm]{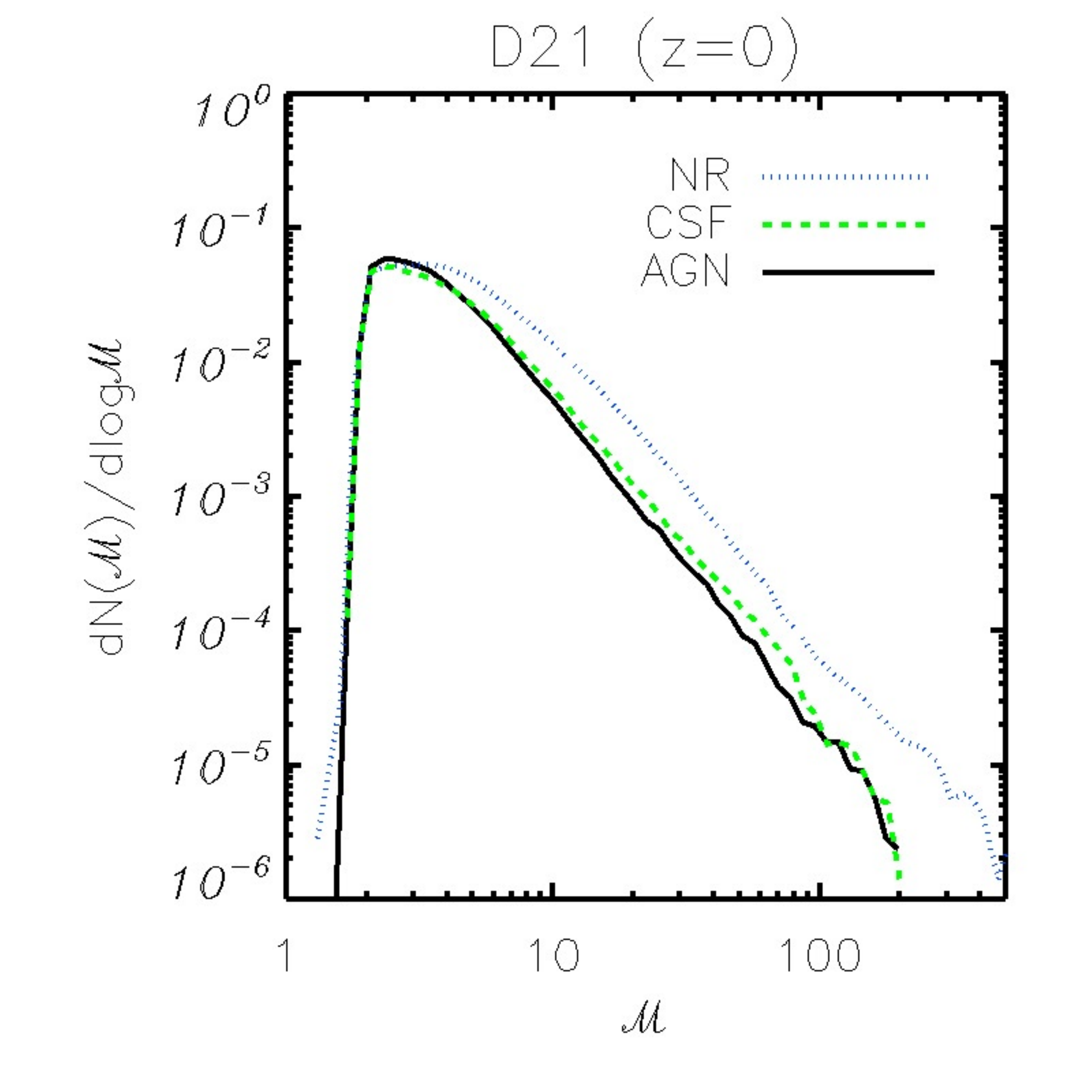}}
\caption{{\it Left panel:} Volume-averaged radial profiles of the mean Mach number for the main central cluster of region D21 as obtained from the {\tt \nr}, {\tt \csf} and {\tt \agn} simulations at $z=0$.  {\it Right panel:} Shock cell distribution function within the region D21 as obtained from the {\tt \nr}, {\tt \csf} and {\tt \agn} simulations at $z=0$.}
\label{fig:distri_physics}
\end{figure*}

In this Appendix we explore the effect of different prescriptions for the physics of baryons on the shock cell distribution. In particular, we compare the distribution of Mach numbers at $z=0$ for the region associated to the most massive cluster in our sample, i.e., region D21, as obtained from the {\tt \nr}, {\tt \csf} and {\tt \agn} simulations.

Figures  \ref{fig:d21_physics_big_region} and \ref{fig:d21_physics} show, respectively,  2-D zoom projections of the gas overdensity  and the Mach number distribution  within the whole considered volume and within the virial radius of cluster  D21 in the {\tt \nr}, {\tt \csf} and {\tt \agn} simulations at $z=0$. Independently of the considered region, although the density distributions of all three runs are quite similar, both radiative runs show a larger number of small density peaks which are mainly associated to  radiative cooling and star formation processes. However, the action of AGN feedback dilutes some of these clumps and generates a much puffier density distribution than in the {\tt \csf} run \citep[see also][]{Planelles_2014}. This effect if more clearly distinguishable in the upper panel of Fig.~\ref{fig:d21_physics}.

As already anticipated in Section \ref{sec:shock_distribution}, Fig.~\ref{fig:d21_physics_big_region} shows that the large-scale shock cell distribution obtained for region D21 is very similar in all three runs. This behaviour was expected, since shocks in cluster outskirts are mainly driven by gravitational effects associated to cosmic structure formation processes. However, as shown in the lower panels of  Fig.~\ref{fig:d21_physics}, some differences in the distribution of shocks obtained in the {\tt \nr}, {\tt \csf} and {\tt \agn} simulations are detectable within the cluster virial region \citep[e.g.][]{Vazza_2013, Schaal_2016}. Although in this particular case these differences are relatively minor, they result from the different radiative and feedback processes included in our simulations.
This behavior is also displayed in the Mach number radial profiles shown in the left panel of Fig.~\ref{fig:distri_physics}.

For the sake of completeness, the right panel of Fig.~\ref{fig:distri_physics} shows the shock cell distribution function within a volume of side length equal $8\times R_{vir}$ around the centre of cluster D21  as obtained in the {\tt \nr}, {\tt \csf} and {\tt \agn} simulations at $z=0$. As already discussed in previous studies \citep[e.g.][]{Vazza_2013}, we find a small shortage of shocks in the {\tt \agn} simulation, especially within $4\mincir\mathcal{M}\mincir60$. This difference is particularly significant between the  {\tt \agn} and the {\tt \nr} simulations: on average, AGN feedback tends to increase the mean ICM temperature, thus affecting the sound speed in the medium and, therefore, decreasing the corresponding strength of shocks. This discrepancy is similarly present when comparing the {\tt \csf} and the {\tt \nr} simulations. Overall, these results suggest that the inclusion of radiative cooling, which removes some gas from the diffuse accretion, plays the main role, while the effect of changing the feedback prescription, namely including or not AGN feedback, is rather minor.

\bsp	
\end{document}